\documentclass{article}
\usepackage{amssymb,amsmath,amsthm,amsbsy,xcolor,graphicx,hyphenat}
\usepackage{titling,float}

\usepackage{geometry}
\numberwithin{equation}{section}

\newcommand{\RR}{{\mathbb{R}}}
\newcommand{\CC}{{\mathbb{C}}}

\newcommand{\pa}{\partial}
\newcommand{\ii}{{\rm i}}
\newcommand{\dd}{{\rm d}}
\newcommand{\Tr}{\mathrm{Tr}}

\newcommand{\bra}[1]{\langle #1 |}
\newcommand{\ket}[1]{| #1 \rangle}
\newcommand{\braket}[2]{\langle #1 | #2 \rangle}
\newcommand{\bP}{\boldsymbol{P}}
\newcommand{\bS}{\boldsymbol{S}}
\newcommand{\bL}{\boldsymbol{L}}
\newcommand{\hx}{\hat{x}}
\newcommand{\bhx}{\hat{\boldsymbol{x}}}

\newcommand{\btau}{\boldsymbol{\tau}}
\newcommand{\bsigma}{\boldsymbol{\sigma}}
\newcommand{\bomega}{\boldsymbol{\omega}}
\newcommand{\bpi}{\boldsymbol{\pi}}
\newcommand{\bx}{\boldsymbol{x}}
\newcommand{\bX}{\boldsymbol{X}}
\newcommand{\tautau}{\boldsymbol{\tau}_1\boldsymbol{\tau}_2}
\newcommand{\sigsig}{\boldsymbol{\sigma}_1\boldsymbol{\sigma}_2}
\newcommand{\Ls}{\boldsymbol{L}\cdot\boldsymbol{\sigma}}
\newcommand{\nl}{\nonumber\\&\phantom{=}}

\begin{document}

\title{Nucleon-nucleon potential from skyrmion dipole interactions: erratum}
%\author{Derek Harland\footnote{email address: d.g.harland@leeds.ac.uk}\,\, and Chris Halcrow\footnote{email address: c.j.halcrow@leeds.ac.uk}\\University of Leeds}
	\author{Chris Halcrow and Derek Harland\\
	School of Mathematics, University of Leeds}
\date{9th August 2022}

\maketitle

Since publication we have noticed two errors in our calculation.  The first is in eq.\ (3.5) for the charge distribution of a dipole, which is missing one term and should read:
\begin{equation}
\label{T corrected}
T:= \left( \left(\boldsymbol{c} -\tfrac{1}{2}|\dot{\boldsymbol{X} }|^2 \boldsymbol{c} - \tfrac{1}{2} (\dot{\boldsymbol{X}}\cdot \boldsymbol{c}) \dot{\boldsymbol{X}} \right) \cdot \nabla + \dot{\boldsymbol{X}}\cdot \boldsymbol{c}\times \boldsymbol{\omega}+\ddot{\boldsymbol{X}}\cdot \boldsymbol{c}  \right).
\end{equation}
This in turn means that eqs.\ (3.17)--(3.20) should read:
\begin{align}
A_{ab;ij}&=\varepsilon_{ajc}(-\delta_{ib}\nabla_c e^{-m r}/r-\tfrac12\nabla_{ibc}e^{-m r}/m) \label{Anew} \\ 
B_{ab;ij}&= -\varepsilon_{aic}\varepsilon_{bjd}\nabla_{cd}e^{-m r}/m \label{Bnew}\\
C_{ab;ij}&=\tfrac12\delta_{ij}\nabla_{ab}e^{-m r}/r -\tfrac14\nabla_{abij}e^{-m r}/m \\& \quad
-\tfrac38(\delta_{jb}\nabla_{ia}
+\delta_{ja}\nabla_{ib}
+\delta_{ib}\nabla_{ja}
+\delta_{ia}\nabla_{jb})e^{-m r}/r
  \nonumber \\
D_{ab}&=\nabla_{ab}e^{-m r}/r. \label{Dnew}
\end{align}
The second error is in eq.\ (4.3), which is missing two terms and should read:
\begin{multline}
H =
- \frac{\hbar^2}{2} \triangle_g + V
+ \frac{\hbar^2}{2} E_\kappa g^{\kappa\lambda}\delta g_{\lambda\mu}g^{\mu\nu}E_\nu
- \frac{\hbar^2}{2} E_\kappa g^{\kappa\lambda}\delta g_{\lambda\mu}g^{\mu\nu}\delta g_{\nu\rho}g^{\rho\sigma}E_\sigma \\
+\frac{\hbar^2}{32} g^{\mu\nu}[E_\mu,g^{\kappa\lambda}\delta g_{\lambda \kappa}][E_\nu,g^{\rho\sigma}\delta g_{\sigma\rho}] 
\textcolor{black}{+\frac{\hbar^2}{8}[E_\mu,g^{\mu\nu}[E_\nu,g^{\kappa\lambda}\delta g_{\kappa\lambda}-\frac{1}{2}g^{\kappa\lambda}\delta g_{\lambda \rho}g^{\rho\sigma}\delta g_{\sigma\lambda}]]} \\
\textcolor{black}{-\frac{\hbar^2}{8}[E_\mu,g^{\mu\nu}\delta g_{\nu\lambda}g^{\lambda\kappa}[E_\kappa,g^{\rho\sigma}\delta g_{\rho\sigma}]]}
+ O(\delta g^3).
\end{multline}
These errors mean that the formulae given for the nucleon-nucleon potentials in eqs.\ (5.39)--(5.47) and (E.1)--(E.8) are incorrect.  The corrected versions of (E.1)--(E.8) are appended, and the corrected versions of (5.39)--(5.47) are obtained from these by setting $s=0$.  We have recalibrated our model using the same method as in our original paper but with the corrected potentials.  The revised calibration is
\begin{equation}
F_\pi = 201 \;\text{ MeV} ,\quad e = 3.30 \quad\text{and}\quad m_\pi = 199 \;\text{ MeV}.
\end{equation}
This is not too different from the calibration originally proposed in (5.50).  It leads to $M=2389\;\text{MeV}$, $\Lambda=321\;\text{MeV\;fm}^2$ and $\rho=143\;\text{MeV\;fm}^3$ in place of (5.51).   The revised version of Fig. \ref{fig:results} of the paper is plotted in Fig. \ref{fig:results_new} below.

As can be seen from Fig. \ref{fig:results_new}, the main consequence of these changes is that the isoscalar spin-orbit potential is no longer attractive in the range of interest.  This is mainly due to the correction to eq.\ (3.5); the correction to eq.\ (4.3) makes little difference to the potentials.  Apart from the two spin-orbit potentials, the remaining 6 potentials are in good agreement with the Paris potential.  In a forthcoming paper, we show that correct signs are obtained for the spin-orbit potentials in a more sophisticated approximation based on instanton holonomies \cite{HHinstantons}.

We should point out that the incorrect eq.\ (4.3) was taken from eq.\ (6.15) of \cite{schroers1993}, which we believe should read
\begin{equation}
\rho_1(t,\boldsymbol{x})=-[(\dot{\boldsymbol{u}}\cdot\boldsymbol{p})+(\boldsymbol{u}\cdot\boldsymbol{p}\times\boldsymbol{\alpha})+(1-\tfrac12|\boldsymbol{u}|^2\boldsymbol{p}\cdot\nabla-\tfrac{1}{2}(\boldsymbol{u})\cdot\boldsymbol{p})(\boldsymbol{u}\cdot\boldsymbol{\nabla})]\delta^3(\boldsymbol{x}-\boldsymbol{X}).
\end{equation}
In \cite{schroers1993} the term $(\dot{\boldsymbol{u}}\cdot\boldsymbol{p})$ was neglected.  If we follow through the calculation of \cite{schroers1993}, and eliminate terms involving accelerations by subtracting total time derivatives, we obtain a lagrangian
\begin{align}
\tilde{L}_{\text{int}} &= 2\kappa\big[
-\nabla\cdot\mathcal{O}\nabla
+\boldsymbol{u}\cdot\mathcal{O}(\nabla\times\boldsymbol{\beta})-(\nabla\times\boldsymbol{\alpha})\cdot\mathcal{O}\boldsymbol{v}
+\tfrac{1}{2}(|\boldsymbol{u}|^2+|\boldsymbol{v}|^2)\nabla\cdot\mathcal{O}\nabla
\nonumber\\&\phantom{=2\kappa\big[}
+(\boldsymbol{u}\cdot\mathcal{O}\nabla)(\boldsymbol{v}\cdot\nabla)+(\nabla\cdot\mathcal{O}\boldsymbol{v})( \boldsymbol{u}\cdot\nabla)
-\tfrac{1}{2}(\boldsymbol{u}\cdot\nabla)(\boldsymbol{u}\cdot\mathcal{O}\nabla)-\tfrac{1}{2}(\boldsymbol{v}\cdot\nabla)(\nabla\cdot\mathcal{O}\boldsymbol{v})
\big]\frac{1}{R}\nonumber\\ &
\phantom{=}+\kappa \big[
(\nabla\times\boldsymbol{\alpha})\cdot\mathcal{O}(\nabla\times\boldsymbol{\beta})
-(\nabla\cdot\mathcal{O}\nabla)(\boldsymbol{u}\cdot\nabla)(\boldsymbol{v}\cdot\nabla)
\nonumber\\&\phantom{=+\kappa\big[}
+\boldsymbol{v}\cdot\nabla (\nabla\times\boldsymbol{\alpha}\cdot\mathcal{O}\nabla)
-(\boldsymbol{u}\cdot\nabla)(\nabla\cdot\mathcal{O}(\nabla\times\boldsymbol{\beta}))
\big]R.
\label{correct}
\end{align}
This has similar structure to eq.\ (6.22) of \cite{schroers1993}, but differs in a few places.

One of the results of \cite{schroers1993} was that the lagrangians calculated using the dipole approximation and the product approximation agree.  In eqs.\ (7.25)--(7.29) of the calculation based on the product approximation, total derivatives are added and terms involving acceleration are discarded.  In general, discarding an acceleration from a lagrangian can lead to incorrect results: for example, subtracting the derivative of $\frac{1}{2}m\boldsymbol{x}\cdot\dot{\boldsymbol{x}}$ from the standard lagrangian $\frac12m|\dot{\boldsymbol{x}}|^2-V(\boldsymbol{x})$ and discarding the acceleration eliminates the kinetic term.

If we repeat the calculation in \cite{schroers1993} and omit this step, we obtain a lagrangian that agrees exactly with \eqref{correct}.  So the important result that the dipole and product approximations lead to the same lagrangian for well-separated skyrmions does appear to be correct.  We have also checked by numerical calculation that the product approximation agrees with \eqref{correct} at large separations; in fact this numerical observation was what prompted us to discover the mistake in eq.\ (3.5) of our paper.
\begin{figure}[H]
\begin{center}
\includegraphics[width=0.8\textwidth]{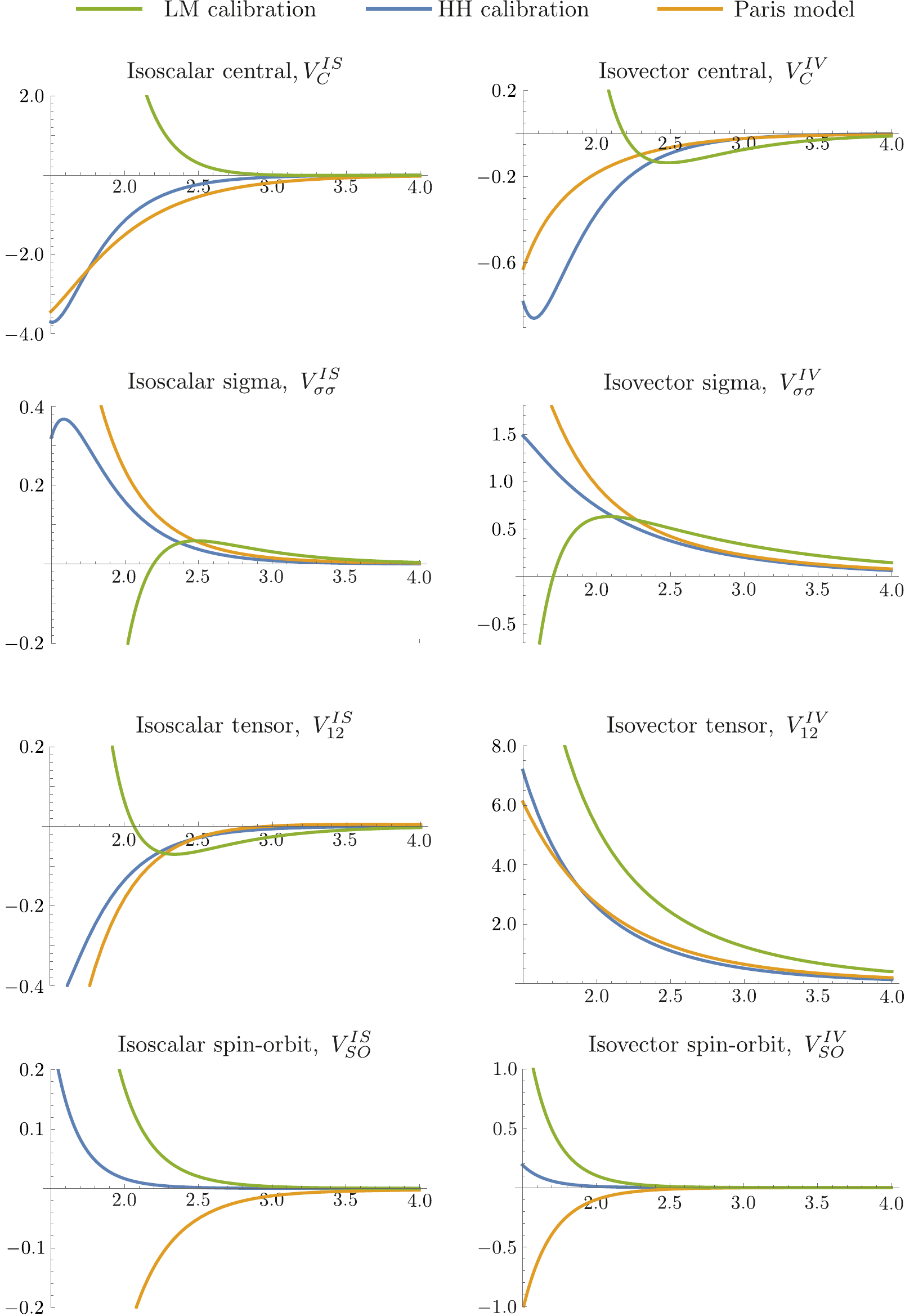}
\end{center}
	\caption{A comparison between the potentials generated from our calculation and the phenomenological Paris potential. All are plots of the potential (MeV) against separation $r$ (fm). This corrects Figure \ref{fig:results} from the main paper below.}
	\label{fig:results_new}
\end{figure}
\addtocounter{figure}{-1}

\section*{Corrected formulae for potentials}
The following correct eqs.\ (E.1)--(E.8) in the original manuscript.  As in those equations, $s:=m_\pi r/\hbar$.
\begin{align}
V_C^{IS}&=\bigg(
-\frac{32 {\Lambda}}{9 {\hbar}^{2} r^{6}}-\frac{64 {\Lambda} s}{9 {\hbar}^{2} r^{6}}-\frac{160 {\Lambda} \,s^{2}}{27 {\hbar}^{2} r^{6}}-\frac{64 {\Lambda} \,s^{3}}{27 {\hbar}^{2} r^{6}}-\frac{16 {\Lambda} \,s^{4}}{27 {\hbar}^{2} r^{6}}\nl
-\frac{8}{27 {\Lambda} \,r^{4}}-\frac{16 s}{27 {\Lambda} \,r^{4}}-\frac{8 s^{2}}{27 {\Lambda} \,r^{4}}-\frac{4 s^{3}}{27 {\Lambda} \,r^{4}}+\frac{17 {\hbar}^{2}}{54 {\Lambda}^{3} r^{2}}+\frac{17 {\hbar}^{2} s^{2}}{108 {\Lambda}^{3} r^{2}}
\bigg)\frac{\rho^2}{e^{2s}}\nl+\bigg(
\frac{800 {\Lambda}^{2}}{27 {\hbar}^{2} r^{8}}+\frac{1600 {\Lambda}^{2} s}{27 {\hbar}^{2} r^{8}}+\frac{160 {\Lambda}^{2} s^{2}}{3 {\hbar}^{2} r^{8}}+\frac{2240 {\Lambda}^{2} s^{3}}{81 {\hbar}^{2} r^{8}}+\frac{80 {\Lambda}^{2} s^{4}}{9 {\hbar}^{2} r^{8}}+\frac{160 {\Lambda}^{2} s^{5}}{81 {\hbar}^{2} r^{8}}\nl
+\frac{80 {\Lambda}^{2} s^{6}}{243 {\hbar}^{2} r^{8}}+\frac{472}{27 r^{6}}+\frac{944 s}{27 r^{6}}+\frac{2504 s^{2}}{81 r^{6}}+\frac{1232 s^{3}}{81 r^{6}}+\frac{380 s^{4}}{81 r^{6}}+\frac{260 s^{5}}{243 r^{6}}\nl
+\frac{10 {\hbar}^{2}}{81 {\Lambda}^{2} r^{4}}+\frac{20 {\hbar}^{2} s}{81 {\Lambda}^{2} r^{4}}-\frac{71 {\hbar}^{2} s^{2}}{81 {\Lambda}^{2} r^{4}}+\frac{5 {\hbar}^{2} s^{3}}{3 {\Lambda}^{2} r^{4}}-\frac{563 {\hbar}^{2} s^{4}}{486 {\Lambda}^{2} r^{4}}
\bigg)\frac{\rho^2}{Me^{2s}}\\
V_{\sigma\sigma}^{IS}&=\bigg(
\frac{8 {\Lambda}}{27 {\hbar}^{2} r^{6}}+\frac{16 {\Lambda} s}{27 {\hbar}^{2} r^{6}}+\frac{40 {\Lambda} \,s^{2}}{81 {\hbar}^{2} r^{6}}+\frac{16 {\Lambda} \,s^{3}}{81 {\hbar}^{2} r^{6}}+\frac{14}{81 {\Lambda} \,r^{4}}+\frac{28 s}{81 {\Lambda} \,r^{4}}+\frac{28 s^{2}}{81 {\Lambda} \,r^{4}}\nl
-\frac{13 {\hbar}^{2}}{648 {\Lambda}^{3} r^{2}}+\frac{13 {\hbar}^{2} s}{324 {\Lambda}^{3} r^{2}}
\bigg)\frac{\rho^2}{e^{2s}}\nl+\bigg(
-\frac{280 {\Lambda}^{2}}{81 {\hbar}^{2} r^{8}}-\frac{560 {\Lambda}^{2} s}{81 {\hbar}^{2} r^{8}}-\frac{56 {\Lambda}^{2} s^{2}}{9 {\hbar}^{2} r^{8}}-\frac{784 {\Lambda}^{2} s^{3}}{243 {\hbar}^{2} r^{8}}-\frac{728 {\Lambda}^{2} s^{4}}{729 {\hbar}^{2} r^{8}}-\frac{112 {\Lambda}^{2} s^{5}}{729 {\hbar}^{2} r^{8}}\nl
-\frac{134}{81 r^{6}}-\frac{268 s}{81 r^{6}}-\frac{2156 s^{2}}{729 r^{6}}-\frac{1096 s^{3}}{729 r^{6}}-\frac{292 s^{4}}{729 r^{6}}\nl
-\frac{265 {\hbar}^{2}}{5832 {\Lambda}^{2} r^{4}}-\frac{265 {\hbar}^{2} s}{2916 {\Lambda}^{2} r^{4}}-\frac{445 {\hbar}^{2} s^{2}}{1944 {\Lambda}^{2} r^{4}}-\frac{661 {\hbar}^{2} s^{3}}{2916 {\Lambda}^{2} r^{4}}
\bigg)\frac{\rho^2}{Me^{2s}}\\
V_{12}^{IS}&=\bigg(
-\frac{8 {\Lambda}}{27 {\hbar}^{2} r^{6}}-\frac{16 {\Lambda} s}{27 {\hbar}^{2} r^{6}}-\frac{32 {\Lambda} \,s^{2}}{81 {\hbar}^{2} r^{6}}-\frac{8 {\Lambda} \,s^{3}}{81 {\hbar}^{2} r^{6}}\nl
-\frac{28}{81 {\Lambda} \,r^{4}}-\frac{35 s}{81 {\Lambda} \,r^{4}}-\frac{14 s^{2}}{81 {\Lambda} \,r^{4}}-\frac{13 {\hbar}^{2}}{648 {\Lambda}^{3} r^{2}}-\frac{13 {\hbar}^{2} s}{648 {\Lambda}^{3} r^{2}}
\bigg)\frac{\rho^2}{e^{2s}}\nl+\bigg(
\frac{224 {\Lambda}^{2}}{81 {\hbar}^{2} r^{8}}+\frac{448 {\Lambda}^{2} s}{81 {\hbar}^{2} r^{8}}+\frac{392 {\Lambda}^{2} s^{2}}{81 {\hbar}^{2} r^{8}}+\frac{560 {\Lambda}^{2} s^{3}}{243 {\hbar}^{2} r^{8}}+\frac{448 {\Lambda}^{2} s^{4}}{729 {\hbar}^{2} r^{8}}+\frac{56 {\Lambda}^{2} s^{5}}{729 {\hbar}^{2} r^{8}}\nl
+\frac{134}{81 r^{6}}+\frac{268 s}{81 r^{6}}+\frac{1900 s^{2}}{729 r^{6}}+\frac{767 s^{3}}{729 r^{6}}+\frac{146 s^{4}}{729 r^{6}}\nl
+\frac{751 {\hbar}^{2}}{2916 {\Lambda}^{2} r^{4}}+\frac{281 {\hbar}^{2} s}{729 {\Lambda}^{2} r^{4}}+\frac{553 {\hbar}^{2} s^{2}}{1944 {\Lambda}^{2} r^{4}}+\frac{661 {\hbar}^{2} s^{3}}{5832 {\Lambda}^{2} r^{4}}
\bigg)\frac{\rho^2}{Me^{2s}}\\
V_{LS}^{IS}&=\bigg(
\frac{16 {\Lambda}^{2}}{27 {\hbar}^{2} r^{8}}+\frac{32 {\Lambda}^{2} s}{27 {\hbar}^{2} r^{8}}+\frac{80 {\Lambda}^{2} s^{2}}{81 {\hbar}^{2} r^{8}}+\frac{32 {\Lambda}^{2} s^{3}}{81 {\hbar}^{2} r^{8}}+\frac{16 {\Lambda}^{2} s^{4}}{243 {\hbar}^{2} r^{8}}+\frac{16}{81 r^{6}}+\frac{32 s}{81 r^{6}}\nl
-\frac{8 s^{2}}{243 r^{6}}-\frac{56 s^{3}}{243 r^{6}}-\frac{101 {\hbar}^{2}}{972 {\Lambda}^{2} r^{4}}+\frac{7 {\hbar}^{2} s}{486 {\Lambda}^{2} r^{4}}+\frac{115 {\hbar}^{2} s^{2}}{972 {\Lambda}^{2} r^{4}}
\bigg)\frac{\rho^2}{Me^{2s}}
\end{align}
\begin{align}
V_C^{IV}&=\bigg(
-\frac{16 {\Lambda}}{27 {\hbar}^{2} r^{6}}-\frac{32 {\Lambda} s}{27 {\hbar}^{2} r^{6}}-\frac{80 {\Lambda} \,s^{2}}{81 {\hbar}^{2} r^{6}}-\frac{32 {\Lambda} \,s^{3}}{81 {\hbar}^{2} r^{6}}-\frac{8 {\Lambda} \,s^{4}}{81 {\hbar}^{2} r^{6}}\nl
-\frac{28}{81 {\Lambda} \,r^{4}}-\frac{56 s}{81 {\Lambda} \,r^{4}}-\frac{28 s^{2}}{81 {\Lambda} \,r^{4}}-\frac{14 s^{3}}{81 {\Lambda} \,r^{4}}-\frac{13 {\hbar}^{2}}{324 {\Lambda}^{3} r^{2}}-\frac{13 {\hbar}^{2} s^{2}}{648 {\Lambda}^{3} r^{2}}
\bigg)\frac{\rho^2}{e^{2s}}\nl+\bigg(
\frac{560 {\Lambda}^{2}}{81 {\hbar}^{2} r^{8}}+\frac{1120 {\Lambda}^{2} s}{81 {\hbar}^{2} r^{8}}+\frac{112 {\Lambda}^{2} s^{2}}{9 {\hbar}^{2} r^{8}}+\frac{1568 {\Lambda}^{2} s^{3}}{243 {\hbar}^{2} r^{8}}+\frac{56 {\Lambda}^{2} s^{4}}{27 {\hbar}^{2} r^{8}}+\frac{112 {\Lambda}^{2} s^{5}}{243 {\hbar}^{2} r^{8}}+\frac{56 {\Lambda}^{2} s^{6}}{729 {\hbar}^{2} r^{8}}\nl
+\frac{268}{81 r^{6}}+\frac{536 s}{81 r^{6}}+\frac{1388 s^{2}}{243 r^{6}}+\frac{632 s^{3}}{243 r^{6}}+\frac{182 s^{4}}{243 r^{6}}+\frac{146 s^{5}}{729 r^{6}}\nl
+\frac{373 {\hbar}^{2}}{972 {\Lambda}^{2} r^{4}}+\frac{373 {\hbar}^{2} s}{486 {\Lambda}^{2} r^{4}}+\frac{505 {\hbar}^{2} s^{2}}{972 {\Lambda}^{2} r^{4}}+\frac{11 {\hbar}^{2} s^{3}}{81 {\Lambda}^{2} r^{4}}+\frac{661 {\hbar}^{2} s^{4}}{5832 {\Lambda}^{2} r^{4}}
\bigg)\frac{\rho^2}{Me^{2s}}\\
V_{\sigma\sigma}^{IV}&=\bigg(
\frac{2 s^{2}}{27 r^{3}}-\frac{2 {\hbar}^{2}}{27 {\Lambda}^{2} r}+\frac{{\hbar}^{2} s}{27 {\Lambda}^{2} r}
\bigg)\frac{\rho}{e^s}\nl+\bigg(
-\frac{{\hbar}^{2} s^{3}}{54 {\Lambda} \,r^{3}}
\bigg)\frac{\rho}{Me^s}\nl+\bigg(
\frac{4 {\Lambda}}{27 {\hbar}^{2} r^{6}}+\frac{8 {\Lambda} s}{27 {\hbar}^{2} r^{6}}+\frac{20 {\Lambda} \,s^{2}}{81 {\hbar}^{2} r^{6}}+\frac{8 {\Lambda} \,s^{3}}{81 {\hbar}^{2} r^{6}}\nl
+\frac{1}{27 {\Lambda} \,r^{4}}+\frac{2 s}{27 {\Lambda} \,r^{4}}+\frac{2 s^{2}}{27 {\Lambda} \,r^{4}}+\frac{43 {\hbar}^{2}}{1296 {\Lambda}^{3} r^{2}}-\frac{43 {\hbar}^{2} s}{648 {\Lambda}^{3} r^{2}}
\bigg)\frac{\rho^2}{e^{2s}}\nl+\bigg(
-\frac{340 {\Lambda}^{2}}{243 {\hbar}^{2} r^{8}}-\frac{680 {\Lambda}^{2} s}{243 {\hbar}^{2} r^{8}}-\frac{68 {\Lambda}^{2} s^{2}}{27 {\hbar}^{2} r^{8}}-\frac{952 {\Lambda}^{2} s^{3}}{729 {\hbar}^{2} r^{8}}-\frac{884 {\Lambda}^{2} s^{4}}{2187 {\hbar}^{2} r^{8}}-\frac{136 {\Lambda}^{2} s^{5}}{2187 {\hbar}^{2} r^{8}}\nl
-\frac{185}{243 r^{6}}-\frac{370 s}{243 r^{6}}-\frac{2978 s^{2}}{2187 r^{6}}-\frac{1516 s^{3}}{2187 r^{6}}-\frac{406 s^{4}}{2187 r^{6}}\nl
-\frac{2621 {\hbar}^{2}}{34992 {\Lambda}^{2} r^{4}}-\frac{2621 {\hbar}^{2} s}{17496 {\Lambda}^{2} r^{4}}-\frac{1973 {\hbar}^{2} s^{2}}{11664 {\Lambda}^{2} r^{4}}-\frac{29 {\hbar}^{2} s^{3}}{17496 {\Lambda}^{2} r^{4}}
\bigg)\frac{\rho^2}{Me^{2s}}\\
V_{12}^{IV}&=\bigg(
\frac{2}{9 r^{3}}+\frac{2 s}{9 r^{3}}+\frac{2 s^{2}}{27 r^{3}}+\frac{{\hbar}^{2}}{27 {\Lambda}^{2} r}+\frac{{\hbar}^{2} s}{27 {\Lambda}^{2} r}
\bigg)\frac{\rho}{e^s}\nl+\bigg(
-\frac{{\hbar}^{2}}{9 {\Lambda} \,r^{3}}-\frac{{\hbar}^{2} s}{9 {\Lambda} \,r^{3}}-\frac{{\hbar}^{2} s^{2}}{18 {\Lambda} \,r^{3}}-\frac{{\hbar}^{2} s^{3}}{54 {\Lambda} \,r^{3}}
\bigg)\frac{\rho}{Me^s}\nl+\bigg(
-\frac{4 {\Lambda}}{27 {\hbar}^{2} r^{6}}-\frac{8 {\Lambda} s}{27 {\hbar}^{2} r^{6}}-\frac{16 {\Lambda} \,s^{2}}{81 {\hbar}^{2} r^{6}}-\frac{4 {\Lambda} \,s^{3}}{81 {\hbar}^{2} r^{6}}\nl
-\frac{2}{27 {\Lambda} \,r^{4}}-\frac{5 s}{54 {\Lambda} \,r^{4}}-\frac{s^{2}}{27 {\Lambda} \,r^{4}}+\frac{43 {\hbar}^{2}}{1296 {\Lambda}^{3} r^{2}}+\frac{43 {\hbar}^{2} s}{1296 {\Lambda}^{3} r^{2}}
\bigg)\frac{\rho^2}{e^{2s}}\nl+\bigg(
\frac{272 {\Lambda}^{2}}{243 {\hbar}^{2} r^{8}}+\frac{544 {\Lambda}^{2} s}{243 {\hbar}^{2} r^{8}}+\frac{476 {\Lambda}^{2} s^{2}}{243 {\hbar}^{2} r^{8}}+\frac{680 {\Lambda}^{2} s^{3}}{729 {\hbar}^{2} r^{8}}+\frac{544 {\Lambda}^{2} s^{4}}{2187 {\hbar}^{2} r^{8}}+\frac{68 {\Lambda}^{2} s^{5}}{2187 {\hbar}^{2} r^{8}}\nl
+\frac{185}{243 r^{6}}+\frac{370 s}{243 r^{6}}+\frac{2626 s^{2}}{2187 r^{6}}+\frac{2125 s^{3}}{4374 r^{6}}+\frac{203 s^{4}}{2187 r^{6}}\nl
+\frac{2135 {\hbar}^{2}}{17496 {\Lambda}^{2} r^{4}}+\frac{1649 {\hbar}^{2} s}{8748 {\Lambda}^{2} r^{4}}+\frac{1001 {\hbar}^{2} s^{2}}{11664 {\Lambda}^{2} r^{4}}+\frac{29 {\hbar}^{2} s^{3}}{34992 {\Lambda}^{2} r^{4}}
\bigg)\frac{\rho^2}{Me^{2s}}\\
V_{LS}^{IV}&=\bigg(
\frac{40 {\Lambda}^{2}}{81 {\hbar}^{2} r^{8}}+\frac{80 {\Lambda}^{2} s}{81 {\hbar}^{2} r^{8}}+\frac{200 {\Lambda}^{2} s^{2}}{243 {\hbar}^{2} r^{8}}+\frac{80 {\Lambda}^{2} s^{3}}{243 {\hbar}^{2} r^{8}}+\frac{40 {\Lambda}^{2} s^{4}}{729 {\hbar}^{2} r^{8}}\nl
-\frac{14}{243 r^{6}}-\frac{28 s}{243 r^{6}}-\frac{128 s^{2}}{729 r^{6}}-\frac{86 s^{3}}{729 r^{6}}\nl
+\frac{197 {\hbar}^{2}}{5832 {\Lambda}^{2} r^{4}}+\frac{89 {\hbar}^{2} s}{2916 {\Lambda}^{2} r^{4}}-\frac{19 {\hbar}^{2} s^{2}}{5832 {\Lambda}^{2} r^{4}}
\bigg)\frac{\rho^2}{Me^{2s}}
\end{align}

\newpage

\title{Nucleon-nucleon potential from skyrmion dipole interactions}
%\author{Derek Harland\footnote{email address: d.g.harland@leeds.ac.uk}\,\, and Chris Halcrow\footnote{email address: c.j.halcrow@leeds.ac.uk}\\University of Leeds}	
	\author{Chris Halcrow and Derek Harland\\
	School of Mathematics, University of Leeds}
\date{24th May 2021}

\maketitle

\begin{abstract}
We derive the nucleon-nucleon interaction from the Skyrme model using second order perturbation theory and the dipole approximation to skyrmion dynamics. Unlike previous derivations, our derivation accounts for the non-trivial kinetic and potential parts of the skyrmion-skyrmion interaction lagrangian and how they couple in the quantum calculation. We derive the eight low energy interaction potentials and compare them with the phenomenological Paris model, finding qualitative agreement in seven cases.
\end{abstract}

%\end{frontmatter}

\section{Introduction}

A nucleon can be modelled as a point particle with spin and isospin degrees of freedom.  The standard way to model the nucleon-nucleon interaction uses a hamiltonian that was first written down in \cite{OM1958}.  The form of this hamiltonian is prescribed by symmetries and it is specified by eight potentials that (in its simplest form) depend only on the separation of the nucleons.  Much effort has been devoted to deriving these potentials from a more fundamental theory. It is well-established that the behaviour of the potentials at large separations is governed by Yukawa's theory of pion exchange \cite{Yukawa1935}.  In contrast, at small separations, QCD effects are important and theorists frequently rely on phenomology, fitting the potentials to experimental data. Several of these semi-phenomenological models have been proposed, such as the Paris and Argonne models \cite{Lacombe:1980dr, Wiringa:1994wb}. While parts of these models are fixed by theory, many parameters are not constrained by theory and must be fitted to data: for example, the Paris model has around sixty unconstrained parameters.  More modern models based on effective field theory have a firmer theoretical foundation, but still involve many unconstrained parameters \cite{Entem:2003ft, Epelbaum:2004fk}.  It seems to be very difficult to derive the nucleon-nucleon potentials from fundamental theory without introducing experimentally-determined parameters.

The Skyrme model is a model of nuclei with roots in QCD that, in its simplest form, has only three unconstrained parameters. It models nucleons using topologically nontrivial field configurations called skyrmions \cite{Skyrme1961}. A skyrmion is a spatially localised soliton that can be described using six degrees of freedom: three for its position, and three for its orientation.

In order to understand the nucleon-nucleon interaction from the Skyrme model one should start by understanding the classical dynamics of two skyrmions.  The two-skyrmion system can be described using a configuration space parametrized by two positions and two orientations, at least when the skyrmions are widely-separated.  To extract the nucleon-nucleon interaction one needs to semiclassically quantise the two-skyrmion dynamics.  So two approximations are needed to derive the nucleon-nucleon interaction: an approximation to the classical dynamics of skyrmions, and a quantisation method.

 The problem of deriving the nucleon-nucleon interaction from the Skyrme model has a long history and is not yet resolved. Early in his development of the model, Skyrme used the product approximation to understand the long range interaction of two skyrmions \cite{Skyrme1962}. Here, the two-skyrmion field is given by the product of the fields of two one-skyrmions. While this is a simple way of generating a two-skyrmion field, there is no reason to trust its validity when the skyrmions are close together. This approximation was used by Vinh Mau et al., who showed that the long-range interactions between skyrmions reproduce the one pion exchange potential of Yukawa \cite{VMLLCL1985}.  Subsequent papers attempted to extract shorter-range parts of the nucleon-nucleon potential, again using the product approximation \cite{JJP1985,NR1986}.  These papers all used what we will call first order perturbation theory to pass from classical skyrmions to quantised nucleons.  Disappointingly, this approach did not result in any medium-range central attraction. This was a major failure: without central attraction there can be no nuclear binding.

The resolution of this problem was found by Walet, Amado and Hosaka \cite{WAH1992, WA1992}, and came in two parts. First, they replaced the product approximation, expanding the classical interaction potential as a Fourier series in relative orientation. The Fourier coefficients were fixed by the Atiyah-Manton approximation, where Skyrme fields are generated using instantons \cite{AM1992}. Secondly, they improved the quantisation technique. The potential energy has a low energy region and  the wavefunction should be concentrated there. To account for this fact the authors used second order perturbation theory, and an attractive central potential was found.  They focused entirely on the potential energy, assuming that the kinetic parts of the interaction were subleading.  A year later, Schroers and Gisiger--Paranjape carefully studied the skyrmion-skyrmion interaction, and found that the kinetic energy is not subleading and can dominate the classical dynamics \cite{schroers1993,GP1994, GP1994_2}. The consequences of this fact for the nucleon-nucleon interaction are explored for the first time in this paper.

Another problem with the Skyrme-derived nucleon-nucleon interaction remained unsolved until recently. The isoscalar spin-orbit potential is essential for describing experimental data from nucleon-nucleon scattering \cite{SM1958, GT1957}, and plays a vital role in the prediction of  magic numbers for larger nuclei \cite{Mayer1948}. Riska and Nyman obtained a satisfactory result for the isovector spin-orbit potential, but their approach resulted in an isoscalar potential with the wrong sign \cite{NR1987, RD1988,Otofuji1988}. Various modifications of the Skyrme model were proposed to correct this result, such as coupling the theory to a dilaton \cite{KE1995} and only including the $\mathcal{L}_6$ term \cite{RS1989} (now known as the BPS model \cite{ANSGW2013}). These did not improve the situation, and the most promising approach was shown by Abada to ignore the dominant contribution \cite{Abada1996}, thereby nullifying the earlier positive conclusion. Although these attempts to fix the spin-orbit problem seem very different, they all share a common feature: they combine the product approximation with first order perturbation theory.

A solution to the spin-orbit problem was found by the authors of this paper \cite{HH2020}. The resolution combines second order perturbation theory with the skyrmion-skyrmion interaction first found by Schroers \cite{schroers1993}. This interaction includes the potential terms as well as the kinetic terms, whose significance was demonstrated by Gisiger and Paranjape \cite{GP1994, GP1994_2}.  In fact, it is a coupling between potential and kinetic terms which provides the most important contribution to the spin-orbit potential. The results of this new method were previously calculated only for the isoscalar spin-orbit potential \cite{HH2020}. In this paper, we present the full nucleon-nucleon interaction arising from this method.  We find a significant improvement over previous attempts to derive the nucleon-nucleon interaction from the Skyrme model.

Our approach is in some ways similar to that of Sugawara and von Hippel \cite{SvH1968}.  In their model, pions can excite nucleons to delta resonances, and this results in a pion-mediated central attraction between nucleons.  Our model is similar but, unlike in \cite{SvH1968}, the nucleon-delta amplitude is determined by theory rather than experiment.  Also, the model of \cite{SvH1968} included an omega meson to account for the short-range parts of the nucleon-nucleon potential, whereas our model only captures long-range parts of the nucleon-nucleon potential.  The Skyrme model does provide a framework to study short-range interactions, and we plan to study these in the future.

Although the results in this paper are for the standard Skyrme model, the methods presented are also valid (perhaps with small modifications) for many modified Skyrme models. These are plentiful \cite{ANSGW2013, GHS2015, He:2015eua, SN2018}. The topic of this paper has a history of mistakes and sign errors in the literature \cite{Abada1996, ASW1993}. For both these reasons, we present our calculation in painstaking detail.

To understand the calculation we first must understand nuclei as quantised skyrmions. This is done in section 2. The dipole approximation, first studied by Schroers and Gisiger--Paranjape for massless pions, is derived in section 3 for massive pions. Section 4 derives a quantum hamiltonian from this classical dipole-dipole lagrangian. We present the calculation of the nucleon-nucleon potential from this hamiltonian in section 5, and draw our conclusions in section 6.  We include four appendices.  These provide further details for our calculations, and present our explicit formulae for the nucleon-nucleon potential (which are too long to include in the main body of the article).

\section{Nuclei as quantised skyrmions}

In this section we review how a quantised skyrmion can be viewed as a nucleon, as was first shown in \cite{ANW1983}.  The Skyrme model is a field theory described by the lagrangian
\begin{equation}
\label{skyrme lagrangian}
\int_{\RR^3} \left(-\frac{F_\pi^2}{16\hbar}\Tr(L_\mu L^\mu) + \frac{\hbar}{32 e^2}\Tr([L_\mu,L_\nu][L^\mu,L^\nu]) - \frac{F_\pi^2m_\pi^2}{8\hbar^3}\Tr(1_2-U)\right)\dd^3x.
\end{equation}
Here $U:\RR^{1,3}\to\mathrm{SU}(2)$, $L_\mu=U^{-1}\pa_\mu U$, $1_2$ is the identity matrix, $F_\pi$ is the pion decay constant, $m_\pi$ is the pion mass and $e$ is a dimensionless coupling constant.  Boundary conditions $U(t,\bx)\to 1_2$ as $|\bx|\to\infty$ are imposed to allow for finite energy, and as a result the model has a topologically conserved quantity, the winding number $B\in\mathbb{Z}$ of $U:S^3\cong\RR^3\cup\{\infty\}\to\mathrm{SU}(2)\cong S^3$.  This winding number $B$ has the physical interpretation of baryon number.

Static solutions of the equations of motion with $B=1$ can be obtained using the hedgehog ansatz:
\begin{equation}
\label{hegehog ansatz}
U_H(\bx) = \exp(-\ii\sigma_j\hx_j f(r)),
\end{equation}
in which $r=|\bx|$, $\hx_j=x_j/r$, $\sigma_j$ are the Pauli matrices and $f:\RR_{\geq0}\to\RR$ is chosen to minimise energy subject to the boundary conditions $f(0)=\pi$ and $f(\infty)=0$.  This hedgehog skyrmion is a soliton whose energy is concentrated at the origin.  Further static $B=1$ solutions can be obtained by acting on the hedgehog with symmetries of the theory, namely translations, rotations, and isorotations (which take the form $U\mapsto QUQ^{-1}$ for $Q\in\mathrm{SU}(2)$).  In fact it suffices to act with translations and isorotations only, as the hedgehog is invariant under a combination of rotations and isorotations.  Thus we obtain solutions of the form
\begin{equation}
\label{moduli space}
U(\bx) = Q U_H(\bx-\bX)Q^{-1}
\end{equation}
parametrised by $\bX\in\RR^3$ and $Q\in\mathrm{SU}(2)$.  The parameters $\bX$ and $Q$ respectively describe the position of the soliton and its orientation.

The family \eqref{moduli space} describes the lowest-energy static configurations in the $B=1$ sector.  To a good approximation, low-energy dynamics in the $B=1$ sector can be described by promoting the parameters $\bX,Q$ to time-dependent functions, i.e.\ by writing $U(t,\bx) = Q(t) U_H(\bx-\bX(t))Q(t)^{-1}$.  The lagrangian that governs this simplified dynamics is
\begin{equation}
L = \frac{M}{2}|\dot{\bX}|^2 + \frac{\Lambda}{2}|\bomega|^2,
\end{equation}
in which $M$ and $\Lambda$ are constants which represent the classical mass and moment of interia of the $B=1$ skyrmion, and
\begin{equation}
\label{omega definition}
-\ii\bomega\cdot\bsigma = 2Q^{-1}\dot{Q}\, ,
\end{equation}
where $\bomega$ is interpreted as the angular velocity of the skyrmion. The equations of motion are that of a free spinning top. Hence, for small kinetic energies, the skyrmion simply moves with constant linear and angular velocities.

In order to make contact with nuclear physics we must quantise the low-energy dynamics of a skyrmion.  The quantum mechanical hamiltonian is
\begin{equation}
H = \frac{1}{2M}|\bP|^2 + \frac{\hbar^2}{2\Lambda}|\bS|^2,
\end{equation}
in which $P_j=-\ii\hbar\pa/\pa X_j$ and
\begin{equation}
-\ii S_j\psi(\bX,Q) = \frac{\dd}{\dd \epsilon}\bigg|_{\epsilon=0} \psi(\bX,Qe^{-\ii\epsilon \sigma_j/2}).
\end{equation}
The operators $S_j$ satisfy $[S_i,S_j]=\ii\epsilon_{ijk}S_k$ and are interpreted physically as spin operators.  In order to understand this hamiltonian we first diagonalise the operator $|\bS|^2=S_iS_i$.  The eigenvalues of this operator are known to be of the form $(n^2-1)/4$ for integers $n\geq 1$.  The corresponding eigenspaces are spanned by wavefunctions $\psi(Q) = \rho^n_{\alpha\beta}(Q)$, where $\rho:\mathrm{SU}(2)\to\mathrm{GL}(n,\CC)$ is the $n$-dimensional irreducible representation of $\mathrm{SU}(2)$ and $\rho^n_{\alpha\beta}$ are its matrix entries (with $1\leq\alpha,\beta\leq n$).  Thus the eigenspace $\mathcal{H}_n$ with eigenvalue $(n^2-1)/4$ is naturally isomorphic to $\CC^n\otimes\CC^n$.  The wavefunction is required to satisfy the Finkelstein-Rubinstein constraint $\psi(\bX,-Q)=-\psi(\bX,Q)$, and as a result only the eigenspaces with $n$ even (corresponding to odd-spin representations) are relevant to the quantum mechanics of the skyrmion \cite{FR1968}.  Thus wavefunctions in the lowest energy eigenspace are functions
\begin{equation}
\psi:\RR^{1,3}\to\mathcal{H}_2\cong \CC^2\otimes\CC^2.
\end{equation}
These describe nucleons: the first factor of $\CC^2$ corresponds to a spin doublet, and the second to an isospin doublet.  The next-lowest eigenspace corresponds to functions
\begin{equation}
\psi:\RR^{1,3}\to\mathcal{H}_4\cong \CC^4\otimes\CC^4
\end{equation}
and describes delta resonances.

Before moving on to investigate the dynamics of two skyrmions we pause to describe some operators acting on the 1-skyrmion Hilbert space that will be relevant to later calculations.  Let $R_{ij}(Q)$ be defined by
\begin{equation}
\label{R definition}
R_{ij}(Q) = \frac12\Tr(\sigma_i Q \sigma_j Q^{-1}),\quad 1\leq i,j\leq 3.
\end{equation}
In other words, $R_{ij}$ are the matrix entries of the adjoint representation of $\mathrm{SU}(2)$.  These act on skyrmion wavefunctions $\psi$ by multiplication.  If $\psi\in \mathcal{H}_n$ then $R_{ij}\psi$ can be written
\begin{equation}
\label{Rpsi}
R_{ij}\psi = \sum_m \kappa_j^{mn}\otimes\lambda_i^{mn}\psi,
\end{equation}
where $\kappa_j^{mn}$ and $\lambda_i^{mn}$ are $m\times n$ matrices of Clebsch-Gordon coefficients and the sum is over $m=2,4$ in the case $n=2$ and over $m=n-2,n,n+2$ in the cases $n\geq 4$.  Our calculations later will involve the matrices $\kappa_j^{mn}$ for $m,n=2,4$, and these are given explicitly in \ref{app:kappa}.  The matrices $\lambda_j^{mn}$ are identical to $\kappa_j^{mn}$, but denoted by a different symbol for clarity (the $\kappa$'s correspond to spin and the $\lambda$'s to isospin).  We also make note of some identities involving these matrices; these can be derived using the matrices given in the appendix.  These identities describe: commutators with spin operators,
\begin{equation}
[S_i,\kappa_j^{mn}\otimes\lambda_l^{mn}] = \ii\varepsilon_{ijk}\kappa_k^{mn}\otimes\lambda_l^{mn};
\end{equation}
contractions with spin operators and epsilon tensors,
\begin{equation}
\label{kappaS}
\begin{aligned}[]
\varepsilon_{ijk}S_j\kappa_k^{22}\otimes\lambda_l^{22} &=\ii\kappa_i^{22}\otimes\lambda_l^{22} &
\varepsilon_{ijk}\kappa_j^{22}\otimes\lambda_l^{22} S_k &= \ii\kappa_i^{22}\otimes\lambda_l^{22} \\
\varepsilon_{ijk}S_j\kappa_k^{24}\otimes\lambda_l^{24} &=-\frac{\ii}{2}\kappa_i^{24}\otimes\lambda_l^{24} &
\varepsilon_{ijk}\kappa_j^{24}\otimes\lambda_l^{24} S_k &= \frac{5\ii}{2}\kappa_i^{24}\otimes\lambda_l^{24} \\
\varepsilon_{ijk}S_j\kappa_k^{42}\otimes\lambda_l^{42} &=\frac{5\ii}{2}\kappa_i^{42}\otimes\lambda_l^{42} &
\varepsilon_{ijk}\kappa_j^{42}\otimes\lambda_l^{42} S_k &= -\frac{\ii}{2}\kappa_i^{42}\otimes\lambda_l^{42};
\end{aligned}
\end{equation}
and substitutions in terms of Pauli matrices,
\begin{equation}
\label{kappakappa}
\begin{aligned}
\kappa_i^{22} &= -\frac{1}{\sqrt{3}}\sigma_i \\
\kappa_i^{22}\kappa_j^{22} &= \frac{1}{3}\bigg(\delta_{ij}+\ii\varepsilon_{ijk}\sigma_k\bigg) \\
\kappa_i^{24}\kappa_j^{42} &= -\frac{\sqrt{2}}{3}\bigg(\delta_{ij}-\frac{\ii}{2}\varepsilon_{ijk}\sigma_k\bigg).
\end{aligned}
\end{equation}

\section{Dipole-dipole lagrangian}

Having understood the dynamics of a single skyrmion, we now consider the dynamics of two well-separated skyrmions, following Schroers \cite{schroers1993}.  To do so, we first investigate the asymptotic tail of a single skyrmion.  Far from the centre of a hedgehog skyrmion, the Skyrme field $U$ is close to the vacuum.  In terms of pion fields, we may write $U(\bx)=\exp(\ii\bpi(\bx)\cdot\bsigma)$ and the lagrangian \eqref{skyrme lagrangian} takes the form
\begin{equation}
L = \frac{F_\pi^2}{8\hbar}\int \left(\pa_\mu\bpi\cdot\pa^\mu\bpi - \frac{m_\pi^2}{\hbar^2}\bpi\cdot\bpi+O(\bpi^4)\right)\dd^3x.
\end{equation}
Thus, far from the centre of the skyrmion the Skyrme lagrangian reduces to the Klein-Gordon lagrangian.  The asymptotic field of the skyrmion with position $\boldsymbol{X}$ takes the form
\begin{equation}
\bpi(\bx) \sim -C_1\left(\frac{2\hbar}{F_\pi e}\right)^2 \left(\frac{1}{|\bx-\bX|^3}+\frac{m_\pi}{\hbar|\bx-\bX|^2}\right)e^{-\frac{m_\pi}{\hbar}|\bx-\bX|}R(Q)(\bx-\bX),
\end{equation}
where $R(Q)$ is the orientation matrix defined in \eqref{R definition} and $C_1$ is a dimensionless constant that can be computed numerically.  We can compare this with the field induced by a dipole with dipole moment $\boldsymbol{c}$:
\begin{equation}
\boldsymbol{c}\cdot\nabla\left(\frac{1}{4\pi r}e^{-\frac{m_\pi r}{\hbar}}\right) = -\left(\frac{1}{r^3}+\frac{m_\pi}{\hbar r^2}\right)e^{-\frac{m_\pi r}{\hbar}}\frac{\boldsymbol{c}\cdot\bx}{4\pi}.
\end{equation}
We see that the $j$th pion field is a dipole with dipole moment $c_i=4\pi(\frac{2\hbar}{F_\pi e})^2 C_1R_{ji}$.

We have learned that, far from its center, a skyrmion resembles a triplet of orthogonal dipoles. As such, we can use the theory of relativistic dipoles to describe the interaction of well separated skyrmions. A single dipole with moment $\boldsymbol{c}$ and position $\boldsymbol{X}(t)$ which is moving slowly with velocity $\dot{\boldsymbol{X}}$ and rotating with angular velocity $\boldsymbol{\omega}$ has charge distribution \cite{schroers1993}
\begin{align} \label{dipolecharge}
\rho_\text{d}(\boldsymbol{x},\boldsymbol{X}(t)) &= -T \delta^{(3)}(\boldsymbol{x} - \boldsymbol{X}(t)) \\
%&\equiv -\left( \left(\boldsymbol{c} -\tfrac{1}{2}\dot{\boldsymbol{X} }^2 \boldsymbol{c} - \tfrac{1}{2} \dot{\boldsymbol{X}}\cdot \boldsymbol{c} \dot{\boldsymbol{X}} \right) \cdot \nabla + \dot{\boldsymbol{X}}\cdot \boldsymbol{c}\times \boldsymbol{\omega}  \right)\delta^{(3)}(\boldsymbol{x} - \boldsymbol{X}(t)) \, .
T&:= \left( \left(\boldsymbol{c} -\tfrac{1}{2}|\dot{\boldsymbol{X} }|^2 \boldsymbol{c} - \tfrac{1}{2} (\dot{\boldsymbol{X}}\cdot \boldsymbol{c}) \dot{\boldsymbol{X}} \right) \cdot \nabla + \dot{\boldsymbol{X}}\cdot \boldsymbol{c}\times \boldsymbol{\omega}  \right).
\end{align}
Here and throughout the calculation we have neglected all terms with more than two time derivatives. The potential due to this dipole satisfies
\begin{align}
\ddot{\phi}_\text{d} - (\Delta - m^2 )\phi_\text{d} = - T \delta^{(3)}(\boldsymbol{x} - \boldsymbol{X}(t) ) \, ,
\end{align}
where $m = m_\pi / \hbar$ is the reduced pion mass. We formally invert this equation and expand in the slow moving approximation
\begin{align} \label{dipole potential}
\phi_\text{d} &= \left((\Delta - m^2)-\partial_t^2 \right)^{-1} T \delta^{(3)}(\boldsymbol{x} - \boldsymbol{X}(t) ) \nonumber \\
&=  T G_m + \frac{d^2}{dt^2}T(\Delta - m^2)^{-1}G_m + ...
\end{align}
where $G_m$ is the Greens function for the Yukawa interaction, given by
\begin{equation}
G_m = -\frac{1}{4\pi r}e^{-m r} \, .
\end{equation}
To find the potential we are left to evaluate $F_m := (\Delta - m^2)^{-1}G_m$. We do this by solving
\begin{equation}
G_m =   (\Delta - m^2)F_m
\end{equation}
whose unique decaying solution is
\begin{equation}
F_m = \frac{1}{8\pi m}e^{-mr} \, .
\end{equation}
It is worth considering the massless limit, which was studied earlier in \cite{schroers1993}.  In that case, \eqref{dipole potential} was solved using special properties of the Laplacian, resulting in
\begin{equation}
F_0 := \Delta^{-1} G_0 = -\frac{r}{8\pi}.
\end{equation}
The expansion of our solution about $m=0$ is
\begin{equation}
\lim_{m\to 0} F_m = \frac{1}{8\pi m} - \frac{r}{8\pi} + O(m).
\end{equation}
Although this diverges as $m\to0$, the lagrangian only depends on derivatives of $F_m$ and the divergent term is constant.  Hence our result agrees with \cite{schroers1993} in the limit $m\to 0$.

Combining all these results and inserting the operator $T$, the potential due to \eqref{dipolecharge} is
\begin{align} \label{potential}
\phi_\text{d}  \approx \,&-\frac{1}{4\pi}  \left( \left(\boldsymbol{c} -\tfrac{1}{2}\dot{\boldsymbol{X} }^2 \boldsymbol{c} - \tfrac{1}{2} \dot{\boldsymbol{X}}\cdot \boldsymbol{c} \dot{\boldsymbol{X}} \right) \cdot \nabla + \dot{\boldsymbol{X}}\cdot \boldsymbol{c}\times \boldsymbol{\omega}  \right)\left( \frac{e^{-m|\boldsymbol{x}-\boldsymbol{X}|}}{|\boldsymbol{x}-\boldsymbol{X}|} \right)  \nonumber\\
&+ \frac{1 }{8\pi m}\frac{d^2}{dt^2} \boldsymbol{c} \cdot \nabla \left(  e^{-m|\boldsymbol{x} - \boldsymbol{X} |} \right) \, .
\end{align}
Once again, we have neglected terms with more than two time derivatives.  Note that this is not the result one finds by simply replacing the massless Greens function for the massive Greens function in the expression for $\phi_d$ from Schroers.

A skyrmion is described by three orthogonal dipoles. So the charge density of, and the potential due to, the skyrmion is simply the sum of those of the dipoles. Let the two skyrmions be labeled by $1$ and $2$, each having their own positions $\boldsymbol{X}_1, \boldsymbol{X}_2$ and angular velocities $\boldsymbol{\omega}_1, \boldsymbol{\omega}_2$. The interaction lagrangian is given by
\begin{equation}
L = \frac{1}{2} \left( \int \phi_1 \rho_2 + \phi_2 \rho_1 \right) \dd^3x\,  .
\end{equation}
This can be evaluated using \eqref{dipolecharge} and \eqref{potential}. Once again we neglect terms with more than two derivatives, and the resulting expression is similar to equation (6.19) in \cite{schroers1993}.  The result can be expressed in terms of $\bx:=\bX_1-\bX_2$, $r = |\bx|$, $q:=Q_1^{-1}Q_2$, and the parameter
\begin{equation}
\rho:=\frac{8\pi \hbar^3C_1^2}{e^4F_\pi^2}.
\end{equation}
The lagrangian obtained is
\begin{multline}
\label{skyrmion-skyrmion lagrangian}
L = \frac{M}{4}\dot{x}^i\dot{x}^i + \frac{\Lambda}{2}\omega_1^i\omega_1^i + \frac{\Lambda}{2}\omega_2^i\omega_2^i \\
+ \rho C_{ij}\dot{x}^i\dot{x}^j + \rho A^1_{ij}\dot{x}^i\omega_1^j + \rho A^2_{ij}\dot{x}^i\omega_2^j + \rho B_{ij}\omega_1^i\omega_2^j- 2\rho D,
\end{multline}
where $A^1_{ij}=A_{ab;ij}R_{ab}$, $A^2_{ij}=A_{ba;ij}R_{ab}$, $B_{ij}=B_{ab;ij}R_{ab}$, $C_{ij}=C_{ab;ij}R_{ab}$, $D=D_{ab}R_{ab}$, and
\begin{align}
A_{ab;ij}&=\varepsilon_{ajc}(\delta_{ic}\nabla_b e^{-m r}/r-\tfrac12\nabla_{ibc}e^{-m r}/m) \label{A} \\ 
B_{ab;ij}&= -\varepsilon_{aic}\varepsilon_{bjd}\nabla_{cd}e^{-m r}/m \label{B}\\
C_{ab;ij}&=\tfrac12\delta_{ij}\nabla_{ab}e^{-m r}/r -\tfrac14\nabla_{abij}e^{-m r}/m \\& \quad
+\tfrac18(\delta_{jb}\nabla_{ia}
+\delta_{ja}\nabla_{ib}
+\delta_{ib}\nabla_{ja}
+\delta_{ia}\nabla_{jb})e^{-m r}/r
  \nonumber \\
D_{ab}&=\nabla_{ab}e^{-m r}/r. \label{D}
\end{align}
Later, we will consider the massless ($m=0$) limit. Hence it is helpful to record the lagrangian in this limit, originally derived in \cite{schroers1993}. It has the same structure as \eqref{skyrmion-skyrmion lagrangian} but with
\begin{align}
	A_{ab;ij}&=\varepsilon_{ajc}(\delta_{ic}\nabla_b 1/r +\tfrac12\nabla_{ibc}\,r)  \\ 
	B_{ab;ij}&= \varepsilon_{aic}\varepsilon_{bjd}\nabla_{cd} \,r  \\
	C_{ab;ij}&=\tfrac12\delta_{ij}\nabla_{ab}1/r +\tfrac14\nabla_{abij}\,r 
	+\tfrac18(\delta_{jb}\nabla_{ia}
	+\delta_{ja}\nabla_{ib}
	+\delta_{ib}\nabla_{ja}
	+\delta_{ia}\nabla_{jb})1/r  \\
	D_{ab}&=\nabla_{ab}1/r.
\end{align}
When $r$ is large, the lagrangian \eqref{skyrmion-skyrmion lagrangian} is a small perturbation of the free lagrangian for a pair of point particles.  It implicitly defines a metric on the configuration space of two skyrmions, and at large separations this metric is guaranteed to be positive definite (i.e.\ Riemannian).  At small separations there is no guarantee that the metric is positive definite, but that is no cause for concern: this lagrangian was derived using the dipole asymptotics of skyrmions, so the approximation is reliable only for well-separated and slowly-moving skyrmions.

\section{Skyrmion-skyrmion hamiltonian}

Having obtained an approximate lagrangian for two skyrmions, we now calculate the corresponding hamiltonian.

In general, the hamiltonian associated to a lagrangian for a particle moving on a Riemannian manifold with metric $g$ under the influence of a potential $V$ is $\frac{\hbar^2}{2}\triangle_g + V$, with $\triangle_g$ being the Laplace-Beltrami operator for the metric $g$.  If the metric is given in the form $g=g_{\mu\nu}e^\mu e^\nu$, with $e^\mu$ being a frame for the cotangent bundle, the Laplace-Beltrami operator is
\begin{equation}
\label{LB definition}
\triangle_g=-(\det g)^{-1/2}E_\mu (\det g)^{1/2}g^{\mu\nu} E_\nu + f_{\mu\lambda}^\lambda g^{\mu\nu}E_\nu,
\end{equation}
with $E_\mu$ being the dual frame for the tangent bundle (such that $e^{\mu}(E_\nu)=\delta^\mu_\nu$) and $f_{\lambda\mu}^\nu$ structure constants defined by $[E_\lambda,E_\mu]=f_{\lambda\mu}^\nu E_\nu$.  A derivation of this formula is given in \ref{app:beltrami}.  The operator $\triangle_g$ is manifestly self-adjoint with respect to the inner product $\braket{\psi}{\psi}_g=\int \overline{\psi}\psi(\det g)^{1/2}e^1\wedge\ldots\wedge e^n$.

If the metric is perturbed to $g+\delta g$ then the correct hamiltonian is $\frac{\hbar^2}{2}\triangle_{g+\delta g} + V$.  This is self-adjoint with respect to the inner product $\braket{\psi}{\psi}_{g+\delta g}$ but not the inner product $\braket{\psi}{\psi}_{g}$.  If we want our deformed hamiltonian to still be self-adjoint with respect to $\braket{\psi}{\psi}_g$ we should instead choose
\begin{equation}
H = \det (1+g^{-1}\delta g)^{1/4}\left(\frac{\hbar^2}{2}\triangle_{g+\delta g} + V\right)\det (1+g^{-1}\delta g)^{-1/4}.
\end{equation}
This can be expanded as a power series in $\delta g$.  Assuming that $f_{\mu\lambda}^\lambda=0$, the terms up to quadratic order are
\begin{multline}
\label{hamiltonian deltag}
H = \frac{\hbar^2}{2}\triangle_{g} + V + \frac{\hbar^2}{2}E_\kappa g^{\kappa\lambda}\delta g_{\lambda\mu}g^{\mu\nu}E_\nu 
-\frac{\hbar^2}{2}E_\kappa g^{\kappa\lambda}\delta g_{\lambda\mu}g^{\mu\nu}\delta g_{\nu\rho} g^{\rho \sigma}E_\sigma \\
+\frac{\hbar^2}{32} g^{\mu\nu}[E_\mu,g^{\kappa\lambda}\delta g_{\lambda \kappa}][E_\nu,g^{\rho\sigma}\delta g_{\sigma\rho}] + O(\delta g^3).
\end{multline}

We will use equation \eqref{hamiltonian deltag} to calculate the hamiltonian for the lagrangian \eqref{skyrmion-skyrmion lagrangian} as a power series in $\rho$.  The $\rho$-independent kinetic terms determine a metric $g$, and the $\rho$-dependent kinetic terms determine a perturbation $\delta g$.  For the frame $e^\mu$ we choose
\begin{equation}
e^j = \dd x_j,\quad e^{j+3}=\Omega_1^j,\quad e^{j+6}=\Omega_2^{j};\quad j=1,2,3,
\end{equation}
where
\begin{equation}
-\Omega_\alpha^j\ii\sigma_j=\boldsymbol{\Omega}_\alpha=2q_\alpha^{-1}\dd q_\alpha
\end{equation}
matching \eqref{omega definition}.  The dual frame is
\begin{equation}
E_j = \frac{\partial}{\partial x_j}=\frac{\ii}{\hbar} P_j,\quad E_{j+3}=-\ii S^1_j,\quad E_{j+6}=-\ii S^2_j;\quad j=1,2,3.
\end{equation}
It is important that the plus and minus signs in these equations are chosen correctly.  The sign of the spin terms is correct because
\begin{equation}
-\ii \sigma_k e^{k+3}(E_{j+3})=
-\ii\sigma_k\Omega_1^k(-\ii S^1_j)=\boldsymbol{\Omega}_1(-\ii S^1_j) = 2Q_1^{-1}\left.\frac{\dd}{\dd t}\right|_{t=0} Q_1 e^{-\ii t\sigma_j/2} = -\ii\sigma_j
\end{equation}
and, more generally, $e^\mu(E_\nu)=\delta^\mu_\nu$.  From equations \eqref{hamiltonian deltag} and \eqref{skyrmion-skyrmion lagrangian} we find 
\begin{equation}
\label{ss hamiltonian}
H=\frac{\hbar^2}{2\Lambda}|\bS^1|^2 + \frac{\hbar^2}{2\Lambda}|\bS^2|^2+\frac{1}{M}|\bP|^2+H_I,
\end{equation}
where
%\begin{equation}
%\label{ss hamiltonian}
%\begin{aligned}
%H_I &= 2\rho D
%{-\frac{\rho\hbar^2}{2\Lambda^2}}\bigg[S^1_iB_{ij}S^2_j+S^2_jB_{ij}S^1_i\bigg]\\
%&+\frac{\rho\hbar}{M\Lambda}\bigg[P_iA^1_{ij}S^1_j+S^1_jA^1_{ij}P_i+P_iA^2_{ij}S^2_j+S^2_jA^2_{ij}P_i\bigg]
%{-\frac{4\rho}{M^2}}P_iC_{ij}P_j \\
%&+\frac{\rho^2\hbar^2}{2\Lambda^2}\bigg[ S^1_i\left(\frac{1}{\Lambda}B_{ij}B_{kj}+\frac{2}{M}A^1_{ji}A^1_{jk}\right)S^1_k + S^2_i\left(\frac{1}{\Lambda}B_{ji}B_{jk}+\frac{2}{M}A^2_{ji}A^2_{jk}\right)S^2_k\bigg] \\
%&+\frac{\rho^2\hbar^2}{M\Lambda^2}\bigg[S^1_iA^1_{ji}A^2_{jk}S^2_k+S^2_kA^1_{ji}A^2_{jk}S^1_i\bigg] \\
%&-\frac{\rho^2\hbar}{M\Lambda}\bigg[ P_i\left(\frac{1}{\Lambda}A^2_{ij}B_{kj}+\frac{4}{M}C_{ij}A^1_{jk}\right)S^1_k+S^1_k\left(\frac{1}{\Lambda}A^2_{ij}B_{kj}+\frac{4}{M}C_{ij}A^1_{jk}\right)P_i \\
%&\qquad\quad +P_i\left(\frac{1}{\Lambda}A^1_{ij}B_{jk}+\frac{4}{M}C_{ij}A^2_{jk}\right)S^2_k+S^2_k\left(\frac{1}{\Lambda}A^1_{ij}B_{jk}+\frac{4}{M}C_{ij}A^2_{jk}\right)P_i\bigg] \\
%&+\frac{2\rho^2}{M^2}P_i\left(\frac{1}{\Lambda}A^1_{ij}A^1_{kj}+\frac{1}{\Lambda}A^2_{ij}A^2_{kj}+\frac{8}{M}C_{ij}C_{jk}\right)P_k\\
%&+\frac{\hbar^2\rho^2}{M^2\Lambda}C_{ab;ii}C_{cd;jj}\left(\delta_{ac}\delta_{bd}-R_{ad}R_{cb}\right) + \frac{\hbar^2\rho^2}{M^3}\nabla_k C_{ii}\nabla_k C_{jj}+O(\rho^3).
%\end{aligned}
%\end{equation}
\begin{multline}
H_I = 2\rho D -\frac{\rho\hbar^2}{2\Lambda^2}B + \frac{\rho\hbar}{M\Lambda}(P_iA_i+A_i^\dagger P_i) \\
+ \frac{\rho^2\hbar^2}{2\Lambda^3}F + \frac{\rho^2\hbar^2}{M\Lambda^2}A_i^\dagger A_i - \frac{\rho^2\hbar}{M\Lambda^2}(P_i\tilde{A}_i+\tilde{A}_i^\dagger P_i) +O(\rho^3)+O(M^{-2})
\end{multline}
and
\begin{align}
B &= S^1_iB_{ij}S^2_j+S^2_jB_{ij}S^1_i \\
A_i &= A^1_{ij}S^1_j+A^2_{ij}S^2_j \\
F &= S^1_iB_{ij}B_{kj}S^1_k + S^2_iB_{ji}B_{jk}S^2_k \\
\tilde{A}_i &= A^2_{ij}B_{kj}S^1_k + A^1_{ij}B_{jk}S^2_k.
\end{align}

\section{Nucleon-nucleon potential}

In the previous section we determined a hamiltonian that describes two interacting skyrmions.  In this section we will apply perturbation theory to calculate a low-energy effective hamiltonian acting on the nucleon-nucleon sector of the skyrmion-skyrmion Hilbert space.

We begin by recalling some essential ideas from degenerate perturbation theory.  Let $H_0$ be a hamiltonian acting on a Hilbert space with energy eigenvalues $E_0<E_1<E_2<\ldots$ and eigenspaces $\mathcal{H}_N$.  Consider a perturbed hamiltonian of the form
\begin{equation}
H(\epsilon)=H_0+\epsilon H_1,
\end{equation}
with $\epsilon$ small.  When $\epsilon=0$ the $E_0$-eigenspace $\mathcal{H}_0$ is invariant under the action of $H(0)$.  As $\epsilon$ moves away from zero this eigenspace is deformed to a subspace $\mathcal{H}_0(\epsilon)$ which is invariant under $H(\epsilon)$ and which is canonically identified with $\mathcal{H}_0$.  Using this identification $\mathcal{H}_0(\epsilon)\cong\mathcal{H}_0$ one obtains an operator $H_E:\mathcal{H}_0\to\mathcal{H}_0$ which describes the action of $H(\epsilon)$ on $\mathcal{H}_0(\epsilon)$ and can be thought of as a low-energy effective hamiltonian for $H(\epsilon)$.  In \ref{app:perturbation theory} we derive the following perturbative formula for this effective hamiltonian:
\begin{multline}
\label{EH definition}
H_E = E_0 + \epsilon H_1^{00} - \epsilon^2\sum_{N>0} \frac{1}{E_N-E_0}H_1^{0N}H_1^{N0} \\
+\epsilon^3\sum_{M,N\neq0}\frac{1}{(E_N-E_0)(E_M-E_0)}H_1^{0N}H_1^{NM}H_1^{M0} \\
- \frac{\epsilon^3}{2}\sum_{N>0} \frac{1}{(E_N-E_0)^2}(H_1^{0N}H_1^{N0}H_1^{00}+H_1^{00}H_1^{0N}H_1^{N0}) + O(\epsilon^4).
\end{multline}
Here $H_1^{NM}:\mathcal{H}_M\to\mathcal{H}_N$ are the projections of $H_1$ such that $H_1=\sum_{M,N}H_1^{NM}$.  Usually in degenerate perturbation theory one works in a basis in which the first (or second) order term is diagonal, and in that case the formula \eqref{EH definition} reduces to well-known formulae for the perturbed eigenvalues.  The advantage of using equation \eqref{EH definition} is that it does not require one to choose any particular basis.

We will apply the formula \eqref{EH definition} to the skyrmion-skyrmion hamiltonian \eqref{ss hamiltonian}, choosing $H_0=\frac{\hbar^2}{2\Lambda}(|\bS^1|^2+|\bS^2|^2)$ and $\epsilon H_1=\frac{1}{2M}|\bP|^2+H_I$.  This means that we have two deformation parameters: $\rho$ and $M^{-1}$ (recall that $H_I=O(\rho)$).  The eigenspaces of $H_0$ are the spaces $\mathcal{H}_N=\mathcal{H}_m\otimes\mathcal{H}_n$ labelled by $N=(m,n)$, which describe particles of spin $\frac{m-1}{2}$ and $\frac{n-1}{2}$.  The associated eigenvalues of $H_0$ are $E_N=(m^2+n^2-2)\hbar^2/8\Lambda$, and the lowest eigenvalue $E_0$ corresponds to the nucleon-nucleon sector labelled by $(m,n)=(2,2)$.

Equation \eqref{EH definition} implies that
\begin{multline}
\label{EH}
H_E = E_0 + \frac{|\bP|^2}{2M}+H_I^{00} - \sum_{N>0}H_I^{0N}\frac{1}{E_N-E_0}H_I^{N0} \\
+ \frac{1}{2M}\sum_{N>0}\frac{1}{(E_N-E_0)^2}(H_I^{0N}[|\bP|^2,H_I^{N0}] - [|\bP|^2,H_I^{0N}]H_I^{N0})\\ + O(\rho^3) + O(M^{-2}).
\end{multline}
The three terms involving $H_I$ will be referred to as first, second and third order.  Notice that the second order term does not involve $|\bP|^2$ because $|\bP|^2$ commutes with $H_0$ and thus $(|\bP|^2)^{N0}=0$ if $N>0$.  The third order term is simpler than in equation \eqref{EH definition} because we are only working up to order 2 in $\rho$ and order 1 in $M^{-1}$.

A key feature of the formula \eqref{EH} is that the sums over $N$ are finite.  This is a special feature of the dipole approximation, and happens because $H_I$ depends on $Q_1,Q_2$ only through the combination $R_{ab}(q)=R_{ca}(Q_1)R_{cb}(Q_2)$.  By the Clebsch-Gordon rules, multiplying a state $\psi\in\mathcal{H}_2\otimes\mathcal{H}_2$ with $R_{ab}(q)$ results in a state in $\bigoplus_{m,n=2,4}\mathcal{H}_m\otimes\mathcal{H}_n$.  More explicitly, from equation \eqref{Rpsi},
\begin{equation}
\label{R representation}
R_{ab}\,\psi = R_{ca}(Q_1)R_{cb}(Q_2)\,\psi = \sum_{m,n=2,4} \kappa_{1a}^{m2}\otimes\lambda_{1c}^{m2}\otimes\kappa_{2b}^{n2}\otimes\lambda_{2c}^{n2}\,\psi
\end{equation}
(where the first subscript on $\kappa$ and $\lambda$ labels the particle).  Thus in equation \eqref{EH definition}, and in what follows, the notation $\sum_{N>0}$ means a sum over $(m,n)=(2,4),(4,2),(4,4)$.  Similarly, $\sum_N$ will mean a sum over $(m,n)=(2,2),(2,4),(4,2),(4,4)$.

We now consider when the use of perturbation theory in \eqref{EH} is justified.  In general, perturbation theory is considered reliable if the correction to the hamiltonian is small compared with the energy differences $E_N-E_0$.  The smallest energy difference is $3\hbar^2/2\Lambda$.  Assuming that the separation $r>\hbar/m_\pi$, the approximation will be reliable provided that
\begin{equation}
\frac{|\bP|^2}{M},\,\frac{\rho}{r^3},\,\frac{\rho\hbar^2}{r\Lambda^2},\,\frac{\rho\hbar|\bP|}{M\Lambda r^2},\,\frac{\rho|\bP|^2}{M^2r^3}< \frac{\hbar^2}{\Lambda}.
\end{equation}
The following conditions on $|\bP|,r,M$ are sufficient to ensure that this is the case:
\begin{equation} \label{ineqs}
|\bP|^2< \frac{M\hbar^2}{\Lambda},\quad r> \max\left\{ \sqrt[3]{\frac{\rho\Lambda}{\hbar^2}},\frac{\rho}{\Lambda},\frac{\hbar}{m_\pi}\right\},\quad M> \frac{\hbar^2}{\Lambda}.
\end{equation}
The first inequality simply means that the skyrmions are moving slowly.  The third inequality is equivalent to the statement that the dominant contribution to the nucleon energy is the rest mass of a skyrmion, rather than its spin energy.  This is true in all proposed calibrations of the Skyrme model.  The condition that $r>\hbar/m_\pi$ is not very restrictive, because the pion Compton wavelength $\hbar/m_\pi$ is only slightly larger than the proton charge radius.  We will examine the remaining constraints in the next section, when we discuss calibrations.

We wish to compare the hamiltonian \eqref{EH} with the nucleon-nucleon potential.  The latter is constrained by symmetry to be of the form \cite{OM1958}
\begin{align} 
V_{NN} &= V_C^{IS} + V_{\sigma\sigma}^{IS}\sigsig+V_{12}^{IS}S_{12} + \tfrac{1}{\hbar}V_{LS}^{IS}\Ls \nonumber \\
&\quad + \left(V_C^{IV} + V_{\sigma\sigma}^{IV}\sigsig+V_{12}^{IV}S_{12} + \tfrac{1}{\hbar}V_{LS}^{IV}\Ls\right)\tautau.
\label{NN potential}
\end{align}
The potentials are known as the isoscalar or isovector central, sigma, tensor and spin-orbit potentials. Here $\sigma_{1i},\sigma_{2i}$ are the spin Pauli matrices and $\tau_{1i},\tau_{2i}$ are the isospin Pauli matrices, and we have used the shorthand $\tautau = \sum_i\tau_{1i}\tau_{2i}$ and $\sigsig=\sum_i\sigma_{1i}\sigma_{2i}$.  The operator $S_{12}$ is $S_{12}=3(\bsigma_1\cdot\bhx)(\bsigma_2\cdot\bhx)-\bsigma_1\cdot\bsigma_2$.  The total spin is $\bsigma=\bsigma_1+\bsigma_2$, and $\bL=\bx\times \bP$ is the total angular momentum.  The coefficient functions $V_{\ast\ast}^{\ast\ast}$ are functions of $r=|\bx|$, $|\bP|^2$ and $|\bL|^2$ only.   Since the skyrmion-skyrmion system shares the symmetries of the nucleon-nucleon system, the hamiltonian \eqref{EH} is guaranteed to be of this form.

We now proceed to describe the calculation of the terms in \eqref{NN potential} from equation \eqref{EH}.  This is a lengthy calculation, and in order to avoid errors we used two independent methods.  The first method is algebraic in character and exploits the identities \eqref{kappaS} and \eqref{kappakappa}.  The second method is a direct calculation in Mathematica that uses the explicit formulae for the matrices $\kappa_i^{mn}$ given in  \ref{app:kappa}.  Both methods gave the same result in the case $m_\pi=0$.  The result for $m_\pi>0$ was obtained using the second method only.

In the following few subsections we explain the algebraic method for evaluating \eqref{EH}.  We assume that $m_\pi=0$ throughout this calculation.  We work to order 2 in $\rho$ and order 1 in $\frac{1}{M}$: thus all equations are understood to be true only up to terms of $O(\rho^3)$ and $O(M^{-2})$.  Further details of the calculation are given in \ref{app:massless pions}.  At the end of this section we give the final result for massless pions, and briefly explain the Mathematica-based calculation.  The final result for massive pions is more complicated, and is given in \ref{app:pion mass}. 

\subsection{First order terms}

In this subsection we evaluate the first order term in equation \eqref{EH}, i.e.\
\begin{multline}
\label{first order}
H_I^{00} = 2\rho D^{00} -\frac{\rho\hbar^2}{2\Lambda^2}B^{00} + \frac{\rho\hbar}{M\Lambda}(P_iA_i+A_i^\dagger P_i)^{00} \\
+ \frac{\rho^2\hbar^2}{2\Lambda^3}F^{00} + \frac{\rho^2\hbar^2}{M\Lambda^2}(A_i^\dagger A_i)^{00} - \frac{\rho^2\hbar}{M\Lambda^2}(P_i\tilde{A}_i+\tilde{A}_i^\dagger P_i)^{00}.
\end{multline}

The first few terms can be evaluated using the identity
\begin{equation}
\label{R}
R_{ab}^{00}=\kappa_{1a}^{22}\lambda_{1c}^{22}\kappa_{2b}^{22}\lambda_{2c}^{22}=\frac{1}{9}\sigma_{1a}
\sigma_{2b}\btau_{1}\btau_{2},
\end{equation}
which follows from \eqref{R representation} and the first equation in \eqref{kappakappa}.  Using identity \eqref{R} we find
\begin{equation}
D^{00} = D_{ab}R_{ab}^{00} = \frac{1}{9}\nabla_{ab}(1/r)R_{ab}^{00} =  \frac{1}{9r^3}S_{12}\btau_1\btau_2.
\end{equation}
This reproduces the well-known result that the dipole potential for skyrmions induces the one-pion exchange potential between nucleons \cite{VMLLCL1985}. Using identities \eqref{R} and \eqref{kappaS} we find
\begin{align}
B^{00} &= B_{ab;ij}(S^1_i R_{ab}^{00}S^2_j+S^2_jR_{ab}^{00}S^1_i) \\
&= 2\nabla_{ab}r R_{ab}^{00} \\
&= \frac{4}{27r}\bsigma_1\bsigma_2\btau_1\btau_2 - \frac{2}{27r}S_{12}\btau_1\btau_2.
\end{align}
Using identity \eqref{R} and noting that $(S^1_j)^{00}=\frac12\sigma_{1j}$ and $(S^2_j)^{00}=\frac12\sigma_{2j}$ we obtain
\begin{align}
(P_iA_i+A_i^\dagger P_i)^{00} &= \frac{1}{2}\{P_i,A_i+A_i^\dagger\}^{00} + \frac{1}{2}[P_i,A_i-A_i^\dagger]^{00} \\
&= \frac{1}{2}\{P_i,A_{ab;ij}\}\big(\{R_{ab},S^1_j\}^{00}+\{R_{ba},S^2_j\}^{00}\big)
\nonumber\\&\quad 
- \frac{\ii\hbar}{2}\nabla_iA_{ab;ij}\big([R_{ab},S^1_j]^{00}+[R_{ba},S^2_j]^{00}\big) \\
&= \frac{1}{18}\{P_i,A_{ab;ij}\}\delta_{aj}(\sigma_{2b}+\sigma_{1b})\btau_1\btau_2
\nonumber\\&\quad 
+ \frac{\hbar}{18}\nabla_iA_{ab;ij}\varepsilon_{ajk}(\sigma_{2b}\sigma_{1k}+\sigma_{1b}\sigma_{2k})\btau_1\btau_2 \label{5.16}\\
&=\frac{4\hbar}{9r^3}S_{12}\btau_1\btau_2.
\end{align}
Note that the first term in equation \eqref{5.16} evaluates to 0, because $A_{ab;ij}$ is skew-symmetric in $a$ and $j$.  This particular property of the dipole lagrangian means that the isovector spin-orbit potential vanishes at order 1 in $\rho$.

To evaluate the next few terms, we need an identity for $(R_{ab}R_{cd})^{00}=\sum_N R_{ab}^{0N}R_{cd}^{N0}$.  From the identities \eqref{R representation} and \eqref{kappakappa} we obtain
\begin{equation}
R_{ab}^{0N}R_{cd}^{N0} = \begin{cases}
\frac{1}{27}(\delta_{ac}+\ii\epsilon_{ace}\sigma_{1e})(\delta_{bd}+\ii\epsilon_{bdf}\sigma_{2f})(1-\frac23\tautau) & N=(2,2) \\
\frac{2}{27}(\delta_{ac}-\frac{\ii}{2}\epsilon_{ace}\sigma_{1e})(\delta_{bd}+\ii\epsilon_{bdf}\sigma_{2f})(1+\frac13\tautau) & N=(4,2) \\
\frac{2}{27}(\delta_{ac}+\ii\epsilon_{ace}\sigma_{1e})(\delta_{bd}-\frac{\ii}{2}\epsilon_{bdf}\sigma_{2f})(1+\frac13\tautau) & N=(2,4) \\
\frac{4}{27}(\delta_{ac}-\frac{\ii}{2}\epsilon_{ace}\sigma_{1e})(\delta_{bd}-\frac{\ii}{2}\epsilon_{bdf}\sigma_{2f})(1-\frac16\tautau) & N=(4,4).
\end{cases}
\end{equation}
Therefore
\begin{equation}
\label{RR}
\sum_N R_{ab}^{0N}R_{cd}^{N0} = \frac13\delta_{ac}\delta_{bd}+\frac{1}{18}\epsilon_{ace}\epsilon_{bdf}\sigma_{1e}\sigma_{2f}\btau_1\btau_2.
\end{equation}
Using equation \eqref{RR} and the fact that $S^\alpha_j=\frac12\sigma_{\alpha j}$, we obtain
\begin{align}
F^{00} &= B_{ij;ab}B_{kj;cd}\sum_N S^1_iR_{ab}^{0N}R_{cd}^{N0}S^1_k + B_{ji;ab}B_{jk;cd}\sum_N S^2_iR_{ab}^{0N}R_{cd}^{N0}S^2_k \\
&= \frac{4}{3r^2} + \frac{1}{18r^2}(S_{12}+\bsigma_1\bsigma_2)\btau_1\btau_2.
\end{align}
Similarly, we obtain
\begin{align}
(P_i\tilde{A}_i+\tilde{A}_i^\dagger P_i)^{00} %\nonumber \\ 
&= \frac{1}{2}\{P_i,\tilde{A}_i+\tilde{A}_i^\dagger\}^{00} + \frac{1}{2}[P_i,\tilde{A}_i-\tilde{A}_i^\dagger]^{00} \\
%&= + A_{ab;ij}B_{cd;jk}\sum_N(S^1_kR_{dc}^{0N}R_{ba}^{N0}+R_{ba}^{0N}R_{dc}^{N0}S^1_k+S^2_kR_{cd}^{N0}R_{ab}^{0N}+R_{ab}^{0N}R_{cd}^{N0}S^2_k) \nonumber \\ &
%+ \frac{\ii\hbar}{2}\nabla_i(A_{ab;ij} B_{cd;jk})\sum_N (S^1_kR_{dc}^{0N}R_{ba}^{N0}-R_{ba}^{0N}R_{dc}^{N0}S^1_k+S^2_kR_{cd}^{N0}R_{ab}^{0N}- R_{ab}^{0N}R_{cd}^{N0}S^2_k) \nonumber \\
&=\frac{1}{r^4}\left(1-\frac{1}{18}\btau_1\btau_2\right)\Ls + \frac{\hbar}{18r^4}\left(\bsigma_1\bsigma_2-2S_{12}\right)\btau_1\btau_2.
\end{align}
and
\begin{align}
(A_i^\dagger A_i)^{00} 	
%&= A_{ab;ij}A_{cd;ik}(S^1_jR_{ab}^{0N}R_{cd}^{N0}S^1_k + S^1_jR_{ab}^{0N}R_{dc}^{N0}S^2_k + S^2_jR_{ba}^{0N}R_{cd}^{N0}S^1_k+S^2_jR_{ba}^{0N}R_{dc}^{N0}S^2_k) \\
&= \frac{13}{6r^4}-\frac{1}{9r^4}S_{12}+\frac{5}{36r^4}\bsigma_1\bsigma_2 \nonumber\\
&\quad -\frac{1}{2r^4}\btau_1\btau_2+\frac{5}{108 r^4}S_{12}\btau_1\btau_2-\frac{1}{27r^4}\bsigma_1\bsigma_2\btau_1\btau_2.
\end{align}
Collecting everything together, we find
\begin{multline}
H_I^{00} = \left[\rho^2\frac{2\hbar^2}{3\Lambda^3r^2} + \frac{\rho^2}{M}\frac{13\hbar^2}{6\Lambda^2 r^4}\right]
- \frac{\rho^2}{M}\frac{\hbar^2}{2\Lambda^2 r^4}\btau_1\btau_2
- \frac{\rho^2}{M}\frac{\hbar^2}{9\Lambda^2 r^4}S_{12}\\
+\left[\rho\left(\frac{\hbar^2}{27\Lambda^2 r} + \frac{2}{9r^3}\right)+\frac{\rho}{M}\frac{4\hbar^2}{9\Lambda r^3} + \rho^2 \frac{\hbar^2}{36\Lambda^3 r^2} + \frac{\rho^2}{M}\frac{17\hbar^2}{108\Lambda^2 r^4}\right]S_{12}\btau_1\btau_2 \\
+ \frac{\rho^2}{M}\frac{5\hbar^2}{36\Lambda^2 r^4}\bsigma_1\bsigma_2
+ \left[-\rho\frac{2\hbar^2}{27\Lambda^2 r} + \rho^2\frac{\hbar^2}{36\Lambda^3 r^2} - \frac{\rho^2}{M}\frac{5\hbar^2}{54\Lambda^2 r^4}\right]\bsigma_1\bsigma_2\btau_1\btau_2 \\
-\frac{\rho^2}{M}\frac{\hbar}{\Lambda^2 r^4}\Ls + \frac{\rho^2}{M}\frac{\hbar}{18\Lambda^2 r^4}\Ls\btau_1\btau_2.
\end{multline}

\subsection{Second order terms}

In this subsection we evaluate the second order term in equation \eqref{EH}, i.e.\
\begin{align}\label{second order}
-\sum_{N>0}& \frac{1}{E_N-E_0}H_I^{0N}H_I^{N0} = 
-4\rho^2\sum_{N>0} \frac{1}{E_N-E_0}D^{0N}D^{N0}\nonumber\\
&+\frac{\rho^2\hbar^2}{\Lambda^2}\sum_{N>0} \frac{1}{E_N-E_0}\left(D^{0N}B^{N0}+B^{0N}D^{N0}\right)
-\frac{\rho^2\hbar^4}{4\Lambda^4}\sum_{N>0} \frac{1}{E_N-E_0}B^{0N}B^{N0}\nonumber\\
&-\frac{2\rho^2\hbar}{M\Lambda}\sum_{N>0} \frac{1}{E_N-E_0}\left(D^{0N}(P_iA_i+A_i^\dagger P_i)^{N0}+(P_iA_i+A_i^\dagger P_i)^{0N}D^{N0}\right)\nonumber\\
&+\frac{\rho^2\hbar^3}{2M\Lambda^3}\sum_{N>0} \frac{1}{E_N-E_0}\left(B^{0N}(P_iA_i+A_i^\dagger P_i)^{N0}+(P_iA_i+A_i^\dagger P_i)^{0N}B^{N0}\right).
\end{align}
We will describe how to evaluate just a couple of the terms in this expression.  The results for all other terms can be found in \ref{app:massless pions}, and their total appears at the end of this section.

The first term that we will evaluate is
\begin{multline}
\label{evaluating 2BD}
\frac{\rho^2\hbar^2}{\Lambda^2}\sum_{N>0} \frac{1}{E_N-E_0}\left(D^{0N}B^{N0}+B^{0N}D^{N0}\right) \\
= -\nabla_{ab}\frac{1}{r}\,\nabla_{cd}r\,\varepsilon_{cij}\varepsilon_{dkl} 
\sum_{N>0} \frac{1}{E_N-E_0} \Big[R_{ab}^{0N}S^1_iR_{jk}^{N0}S^2_l+R_{ab}^{0N}S^2_kR_{il}^{N0}S^1_j \\
+ S^1_iR_{jk}^{0N}S^2_lR_{ab}^{N0}+S^2_kR_{il}^{0N}S^1_jR_{ab}^{N0}\Big].
\end{multline}
All of the terms in the sum can be expressed using the identities \eqref{kappaS}, \eqref{kappakappa} and \eqref{R representation}.  For example,
\begin{multline}
\frac{1}{E_N-E_0}\,R_{ab}^{0N}\,\varepsilon_{cij}\varepsilon_{dkl}S^1_iR_{jk}^{N0}S^2_l \\
= \begin{cases}
\frac{2\hbar^2}{3\Lambda}\frac{5\ii}{2}\ii\frac{2}{27}(\delta_{ac}-\frac{\ii}{2}\epsilon_{ace}\sigma_{1e})(\delta_{bd}+\ii\epsilon_{bdf}\sigma_{2f})(1+\frac13\tautau) & N=(4,2) \\
\frac{2\hbar^2}{3\Lambda}\ii(-\frac{\ii}{2})\frac{2}{27}(\delta_{ac}+\ii\epsilon_{ace}\sigma_{1e})(\delta_{bd}-\frac{\ii}{2}\epsilon_{bdf}\sigma_{2f})(1+\frac13\tautau) & N=(2,4) \\
\frac{\hbar^2}{3\Lambda}\frac{5\ii}{2}(-\frac{\ii}{2})\frac{4}{27}(\delta_{ac}-\frac{\ii}{2}\epsilon_{ace}\sigma_{1e})(\delta_{bd}-\frac{\ii}{2}\epsilon_{bdf}\sigma_{2f})(1-\frac16\tautau) & N=(4,4).
\end{cases}
\end{multline}
Evaluating the remaining terms in a similar manner results in an identity
\begin{multline}\label{identity 1}
\varepsilon_{cij}\varepsilon_{dkl} 
\sum_{N>0} \frac{1}{E_N-E_0} \Big[R_{ab}^{0N}S^1_iR_{jk}^{N0}S^2_l+R_{ab}^{0N}S^2_kR_{il}^{N0}S^1_j
+ S^1_iR_{jk}^{0N}S^2_lR_{ab}^{N0}+S^2_kR_{il}^{0N}S^1_jR_{ab}^{N0}\Big] \\
= \frac{\Lambda}{\hbar^2}\delta_{ac}\delta_{bd}\left(-\frac{4}{27}-\frac{14}{81}\tautau\right) + 
\frac{\Lambda}{\hbar^2}\varepsilon_{ace}\varepsilon_{bdf}\sigma_{1e}\sigma_{2f}\left(-\frac{7}{27}-\frac{1}{18}\tautau\right).
\end{multline}
Substituting this into equation \eqref{evaluating 2BD} leads to
\begin{multline}
\frac{\rho^2\hbar^2}{\Lambda^2}\sum_{N>0} \frac{1}{E_N-E_0}\left(D^{0N}B^{N0}+B^{0N}D^{N0}\right)\\
%&=\frac{\rho^2\hbar^2}{\Lambda^2} D_{ab}B_{cd;ij}\sum_{N>0} \frac{1}{E_N-E_0}\left(R_{ab}^{0N}\left(S^1_iR_{cd}^{N0}S^2_j+S^2_jR_{cd}^{N0}S^1_i\right)+\left(S^1_iR_{cd}^{0N}S^2_j+S^2_jR_{cd}^{0N}S^1_i\right)R_{ab}^{N0}\right)\nonumber\\
=\frac{\rho^2}{\Lambda r^4}\left[-\frac{8}{27}-\frac{28}{81}\tautau+\frac{1}{81}(14+3\tautau)(\sigsig-2S_{12})\right].
\end{multline}
The other terms in equation \eqref{second order} that do not involve $P_i$ can be evaluated by a similar method.

Now we evaluate the term involving $P_i$, $A_i$ and $D$.  This requires some algebraic rearrangement:
\begin{align}\label{evaluating 2PAD}
-\frac{2\rho^2\hbar}{M\Lambda}&\sum_{N>0} \frac{1}{E_N-E_0}\left(D^{0N}(P_iA_i+A_i^\dagger P_i)^{N0}+(P_iA_i+A_i^\dagger P_i)^{0N}D^{N0}\right) \nonumber\\
=& -\frac{\rho^2\hbar}{M\Lambda}\sum_{N>0}\frac{1}{E_N-E_0}\left\{P_i,(A_i+A_i^\dagger)^{0N}D^{N0}+D^{0N}(A_i+A_i^\dagger)^{N0}\right\} \nonumber\\
&+\frac{\ii\rho^2\hbar^2}{M\Lambda}\sum_{N>0}\frac{1}{E_N-E_0}\left(D^{0N}\nabla_i(A_i-A_i^\dagger)^{N0}+\nabla_i(A_i-A_i^\dagger)^{0N}D^{N0}\right) \nonumber\\
&+\frac{\ii\rho^2\hbar^2}{M\Lambda}\sum_{N>0}\frac{1}{E_N-E_0}\left((A_i+A_i^\dagger)^{0N}\nabla_iD^{N0} - \nabla_i D^{0N}(A_i+A_i^\dagger)^{N0}\right).
\end{align}
Here we have used the identity $[P_i,\cdot]=-\ii\hbar\nabla_i$.  Each of these three terms can be evaluated using methods similar to those described above.  For example, for the first term we use the following identities, whose derivation is similar to that of \eqref{identity 1}:
\begin{align}
\varepsilon_{cij}\sum_{N>0}&\frac{1}{E_N-E_0}\left[R_{ab}^{0N}(S^1_iR_{jd}^{N0}-R_{id}^{N0}S^1_j) +(S^1_iR_{jd}^{0N}-R_{id}^{0N}S^1_j)R_{ab}^{N0}\right] \nonumber \\
&=\frac{\Lambda}{\hbar^2}\delta_{bd}\varepsilon_{ace}\sigma_{1e}\left(\frac{8}{27}+\frac{2}{81}\tautau\right) + \frac{\Lambda}{\hbar^2}\delta_{ac}\varepsilon_{bdf}\sigma_{2f}\left(-\frac{4}{27}-\frac{10}{81}\tautau\right) \\
\varepsilon_{dij}\sum_{N>0}&\frac{1}{E_N-E_0}\left[R_{ab}^{0N}(S^2_iR_{cj}^{N0}-R_{ci}^{N0}S^2_j) + (S^2_iR_{cj}^{0N}-R_{ci}^{0N}S^2_j)R_{ab}^{N0}\right] \nonumber \\
&= \frac{\Lambda}{\hbar^2}\delta_{bd}\varepsilon_{ace}\sigma_{1e}\left(-\frac{4}{27}-\frac{10}{81}\tautau\right) + \delta_{ac}\varepsilon_{bdf}\sigma_{2f}\left(\frac{8}{27} + \frac{2}{81}\tautau\right).
\end{align}
The result is
\begin{multline}
-\frac{\rho^2\hbar}{M\Lambda}\sum_{N>0}\frac{1}{E_N-E_0}\left\{P_i,(A_i+A_i^\dagger)^{0N}D^{N0}+D^{0N}(A_i+A_i^\dagger)^{N0}\right\} \\
= \frac{\rho^2}{M\hbar r^6}\left(-\frac{4}{3}+\frac{4}{9}\tautau\right)\Ls.
\end{multline}
The full result for equation \eqref{evaluating 2PAD} is in \ref{app:massless pions}.  The remaining term in \eqref{second order}, which involves $P_i$, $A$ and $B$, can be calculated by a similar method and is also given in the appendix.

The complete result for equation \eqref{second order} is
\begin{align}
-&\sum_{N>0}\frac{1}{E_N-E_0}H_I^{0N}H_I^{N0} \nonumber\\
&=
\left[-\rho^2\left(\frac{19\hbar^2}{54\Lambda^3r^2}+\frac{8}{27\Lambda r^4}+\frac{32\Lambda}{9\hbar^2r^6}\right) -\frac{\rho^2}{M}\left(\frac{26\hbar^2}{27\Lambda^2r^4}+\frac{32}{9 r^6}\right)\right]\nonumber\\
& +\left[ -\rho^2\left(\frac{13\hbar^2}{324\Lambda^3 r^2}+\frac{28}{81\Lambda r^4}+\frac{16\Lambda}{27\hbar^2 r^6}\right) +\frac{\rho^2}{M}\left(\frac{\hbar^2}{9\Lambda^2 r^4}-\frac{40}{27 r^6}\right) \right]\tautau \nonumber\\
&+\left[ -\rho^2\left(\frac{13\hbar^2}{648\Lambda^3 r^2}+\frac{28}{81\Lambda r^4}+\frac{8\Lambda}{27\hbar^2 r^6}\right) -\frac{\rho^2}{M}\left(\frac{43\hbar^2}{81\Lambda^2 r^4}+\frac{20}{27 r^6}  \right) \right]S_{12}\nonumber \\
&+\left[ -\rho^2\left(\frac{17\hbar^2}{1296\Lambda^3 r^2}+\frac{6}{81\Lambda r^4}+\frac{4\Lambda}{27\hbar^2 r^6}\right) - \frac{\rho^2}{M}\left(\frac{61\hbar^2}{486\Lambda^2 r^4}+\frac{2}{9 r^6}  \right) \right]S_{12}\tautau\nonumber \\
&+\left[ \rho^2\left(-\frac{13\hbar^2}{648\Lambda^3 r^2}+\frac{14}{81\Lambda r^4}+\frac{8\Lambda}{27\hbar^2 r^6}\right) + \frac{\rho^2}{M}\left(\frac{43\hbar^2}{162\Lambda^2 r^4}+\frac{20}{27 r^6}  \right) \right]\sigsig\nonumber \\
&+\left[ \rho^2\left(-\frac{17\hbar^2}{1296\Lambda^3 r^2}+\frac{1}{27\Lambda r^4}+\frac{4\Lambda}{27\hbar^2 r^6}\right) + \frac{\rho^2}{M}\left(\frac{61\hbar^2}{972\Lambda^2 r^4}+\frac{2}{9 r^6}  \right) \right]\sigsig\tautau\nonumber \\
&+\frac{\rho^2}{M}\left[\frac{20\hbar}{27\Lambda^2 r^4}-\frac{4}{3\hbar r^6}\right]\Ls 
+\frac{\rho^2}{M}\left[\frac{5\hbar}{81\Lambda^2 r^6}+\frac{4}{9\hbar r^6}\right]\Ls\tautau.
\end{align}

\subsection{Third order terms}

In this subsection we evaluate the third order term in equation \eqref{EH}.  We rearrange this as follows:
\begin{multline}
\frac{1}{2M}\sum_{N>0}\frac{1}{(E_N-E_0)^2}(H_I^{0N}[|P|^2,H_I^{N0}] - [|P|^2,H_I^{0N}]H_I^{N0}) \\
= \frac{\ii\hbar}{M}\sum_{N>0} \frac{1}{(E_N-E_0)^2}\big\{P_i,\, \nabla_iH_I^{0N}H_I^{N0}-H_I^{0N}\nabla_iH_I^{N0}\big\} \\
+ \frac{\hbar^2}{M}\sum_{N>0} \frac{1}{(E_N-E_0)^2}\nabla_iH_I^{0N}\nabla_iH_I^{N0}.
\end{multline}
Substituting for $H_I$ leads to
\begin{align}
&\frac{1}{2M}\sum_{N>0}\frac{1}{(E_N-E_0)^2}(H_I^{0N}[|P|^2,H_I^{N0}] - [|P|^2,H_I^{0N}]H_I^{N0}) \nonumber\\
&= \frac{4\ii\rho^2\hbar}{M}\sum_{N>0} \frac{1}{(E_N-E_0)^2}\big\{P_i,\, \nabla_iD^{0N}D^{N0}-D^{0N}\nabla_iD^{N0}\big\} \nonumber\\
&\quad -\frac{\ii\rho^2\hbar^3}{M\Lambda^2}\sum_{N>0} \frac{1}{(E_N-E_0)^2}\big\{P_i,\, \nabla_iD_I^{0N}B_I^{N0}-D_I^{0N}\nabla_iB_I^{N0}+\nabla_iB_I^{0N}D_I^{N0}-B_I^{0N}\nabla_iD_I^{N0}\big\} \nonumber\\
&\quad +\frac{\ii\rho^2\hbar^5}{4M\Lambda^4}\sum_{N>0} \frac{1}{(E_N-E_0)^2}\big\{P_i,\, \nabla_iB_I^{0N}B_I^{N0}-B_I^{0N}\nabla_iB_I^{N0}\big\} \nonumber\\
&\quad +\frac{4\rho^2\hbar^2}{M}\sum_{N>0} \frac{1}{(E_N-E_0)^2}\nabla_iD_I^{0N}\nabla_iD_I^{N0} \nonumber\\
&\quad -\frac{\rho^2\hbar^4}{M\Lambda^2}\sum_{N>0} \frac{1}{(E_N-E_0)^2}\big(\nabla_iD_I^{0N}\nabla_iB_I^{N0}+\nabla_iB_I^{0N}\nabla_iD_I^{N0}\big) \nonumber\\
&\quad +\frac{\rho^2\hbar^6}{4M\Lambda^4}\sum_{N>0} \frac{1}{(E_N-E_0)^2}\nabla_iB_I^{0N}\nabla_iB_I^{N0}.
\label{third order}
\end{align}
Each of these six terms can be evaluated by similar methods to those described above, and the resulting expressions are given in \ref{app:massless pions}.  The end result is
\begin{align}
&\frac{1}{2M}\sum_{N>0}\frac{1}{(E_N-E_0)^2}(H_I^{0N}[|P|^2,H_I^{N0}] - [|P|^2,H_I^{0N}]H_I^{N0}) \nonumber\\
&=\frac{\rho^2}{M}\bigg[ \left(\frac{89\hbar^2}{162\Lambda^2r^4}+\frac{88}{27r^6}+\frac{800\Lambda^2}{27\hbar^2r^8}\right) + \left(\frac{103\hbar^2}{972\Lambda^2 r^4} + \frac{148}{81r^6} + \frac{560\Lambda^2}{81\hbar^2r^8}\right)\tautau \nonumber\\
&+ \left(\frac{103\hbar^2}{2916\Lambda^2r^4} + \frac{74}{81r^6}+\frac{224\Lambda^2}{81\hbar^2 r^8}\right)S_{12} + \left(\frac{281\hbar^2}{17496\Lambda^2r^4} + \frac{59}{243r^6}+\frac{272\Lambda^2}{243\hbar^2r^8}\right)S_{12}\tautau\nonumber \\
&+\left(-\frac{103\hbar^2}{5832\Lambda^2r^4} - \frac{74}{81r^6} - \frac{280\Lambda^2}{81\hbar^2r^8}\right)\sigsig + \left( -\frac{281\hbar^2}{34992\Lambda^2r^4}-\frac{59}{243r^6}-\frac{340\Lambda^2}{243\hbar^2r^8}\right)\sigsig\tautau \nonumber\\
&+\left(\frac{7\hbar^2}{972\Lambda^2r^4}+\frac{52}{81r^6}+\frac{16\Lambda^2}{27\hbar^2r^8}\right)\frac{\Ls}{\hbar} + \left(\frac{89\hbar^2}{5832\Lambda^2r^4}+\frac{22}{243r^6}+\frac{40\Lambda^2}{81\hbar^2r^8}\right)\frac{\Ls}{\hbar}\tautau \bigg].
\end{align}

\subsection{Results}
\label{sec:massless pions}

Adding the first, second and third order results together, we find the final expression of our calculation. The isoscalar potentials are
\begin{align}
V_C^{IS} &=\rho^2\frac{\left(-192 \Lambda^4-16 \Lambda^2 \hbar^2r^2+17\hbar^4r^4\right)}{54
	\Lambda^3 \hbar^2 r^6}
+\frac{\rho^2}{M}\frac{2 \left(1200 \Lambda^4-12 \Lambda ^2 \hbar^2 r^2+71 \hbar ^4r^4\right)}{81 \Lambda^2 \hbar^2 r^8} \\
V_{12}^{IS} &= \rho^2\frac{\left(-192 \Lambda ^4-224 \Lambda ^2 \hbar^2 r^2-13 \hbar^4 r^4\right)}{648 \Lambda^3 \hbar^2 r^6} + \frac{\rho^2}{M}\frac{\left(8064 \Lambda^4+504 \Lambda^2 \hbar^2 r^2-1769 \hbar^4 r^4\right)}{2916 \Lambda^2 \hbar^2 r^8} \\
V_{\sigma\sigma}^{IS} &= \rho^2\frac{ \left(192 \Lambda ^4+112 \Lambda^2 \hbar^2 r^2 -13 \hbar^4 r^4\right)}{648 \Lambda^3 \hbar^2 r^6}+\frac{\rho^2}{M}\frac{\left(-20160 \Lambda^4-1008 \Lambda^2 \hbar^2 r^2 +2255 \hbar^4 r^4\right)}{5832 \Lambda^2 \hbar^2 r^8} \\
V_{SO}^{IS} &= \frac{\rho^2}{M} \frac{\left(24 \Lambda^2-35 \hbar^2 r^2\right) \left(24 \Lambda^2+7 \hbar^2 r^2\right)}{972 \Lambda^2 \hbar^2 r^8} \, ,
\end{align}
while the isovector potentials are
\begin{align}
V_C^{IV} &= \rho^2\frac{\left(-192 \Lambda^4-112 \Lambda^2 \hbar^2 r^2-13 \hbar^4 r^4\right)}{324 \Lambda^3 \hbar^2 r^6} + \frac{\rho^2}{M} \frac{\left(6720 \Lambda^4+336 \Lambda^2 \hbar^2 r^2-275 \hbar ^4 r^4\right)}{972 \Lambda^2 \hbar^2 r^8} \\
V_{12}^{IV} &= \rho \frac{6\Lambda^2+\hbar^2 r^2 }{27\Lambda^2 r^3}+\rho^2 \frac{\left(-192 \Lambda^4-96 \Lambda^2 \hbar^2 r^2 +19 \hbar^4 r^4\right)}{1296 \Lambda^3 \hbar^2 r^6}\\
&+\frac{\rho}{M}\frac{4\hbar^2}{9\Lambda r^3}  + \frac{\rho^2}{M} \frac{\left(19584 \Lambda^4+360 \Lambda^2 r^2 \hbar^2  +839 \hbar^4 r^4\right)}{17496 \Lambda^2 \hbar^2 r^8} \\
V_{\sigma\sigma}^{IV} &= -\rho\frac{2\hbar^2}{27\Lambda^2 r} + \rho^2\frac{\left(192 \Lambda^4+48 \Lambda^2 \hbar^2 r^2 +19 \hbar^4 r^4\right)}{1296 \Lambda^3 \hbar^2 r^6}\nonumber\\
&\quad +\frac{\rho^2}{M} \frac{5 \left(-9792 \Lambda^4-144 \Lambda^2 \hbar^2 r^2-265 \hbar^4 r^4\right)}{34992 \Lambda^2 \hbar^2 r^8}  \\
V_{SO}^{IV} &= \frac{\rho^2}{M}\frac{\left(2880 \Lambda^4+3120 \Lambda^2 \hbar^2 r^2+773 \hbar^4 r^4\right)}{5832 \Lambda^2 \hbar^2 r^8}\, .
\end{align}
The equivalent expressions with a non-zero pion mass can be found in \ref{app:pion mass}.

Finally, we explain our second method for evaluating this potential.  This method started with the same expressions \eqref{first order}, \eqref{second order} and \eqref{third order} but differed in the way the terms in these expressions were evaluated.  Using the identity \eqref{R representation}, the operators $R^{MN}_{ab}$ and $S^\alpha_i$ were replaced with the matrices given in \ref{app:kappa}.  The identities listed in \ref{app:pion mass} were then obtained by computing the resulting matrix products in Mathematica, and the results added together.

\subsection{Comparison with the Paris potential}

We now compare the results of our calculation with the successful semi\hyp{}phenomenological model proposed by the Paris group \cite{Lacombe:1980dr}.  To compare models we must choose a calibration by fixing $F_\pi$, $e$ and $m_\pi$. This is equivalent to fixing the energy scale, length scale and pion mass. Once these are chosen, all other constants are fixed by the Skyrme model. One calibration we consider was proposed by Lau and Manton (LM), optimised to reproduce the Carbon-12 energy spectrum \cite{LM2014}. In this case
\begin{equation}
F_\pi = 108 \text{ MeV} , e = 3.93 \text{ and } m_\pi =  149 \text{ MeV}
\end{equation}
which fixes the constants
\begin{equation}
M = 1096 \text{ MeV} , \Lambda  = 332 \text{ MeV } \text{fm}^2 \text{ and } \rho =  229 \text{ MeV } \text{fm}^3 \, .
\end{equation}
We will also consider a new calibration, optimised to reproduce the Paris potentials. To find this, we consider the sum of the $L^2$ norms of the differences between ours and the Paris potentials for $r = 1.5 - 2.5 \text{ fm}$. This is a function of $F_\pi, e$ and $m_\pi$ and we minimise the function using a numerical gradient flow. We find that the optimal calibration is 
\begin{equation}
F_\pi = 165 \text{ MeV} , e = 3.75 \text{ and } m_\pi =  216 \text{ MeV}
\end{equation}
which fixes the constants
\begin{equation}
M = 1752 \text{ MeV} , \Lambda  = 252 \text{ MeV } \text{fm}^2 \text{ and } \rho =  124 \text{ MeV } \text{fm}^3 \, .
\end{equation} 
We call this the HH calibration. This calibration gives values of $F_\pi$ and $m_\pi$ reasonably close to their physical values. This is expected, since we are dealing with pionic physics. Unfortunately the skyrmion mass $M$ is much too large. This is a common problem when one tries to describe the physics of the nucleon sector. Meier and Walliser proposed that one-loop corrections can significantly reduce the mass \cite{MW1996}, although including this correction is difficult and will affect the interaction potentials we have derived.

Our calculation depended on two approximations: perturbation theory and the dipole approximation to skyrmion dynamics.  Our use of perturbation theory is justified only when the inequalities \eqref{ineqs} hold. In the Lau-Manton calibration the tightest constraint is
\begin{equation} \label{ineqbreak}
r >\frac{\hbar}{m_\pi} = 1.33 \text{ fm} \, ,
\end{equation}
while in the new calibration it is
\begin{equation}
r > \left( \frac{\rho \Lambda}{\hbar^2} \right)^{\tfrac{1}{3}}  = 0.93 \text{ fm} \, .
\end{equation}
It is harder to quantify when the dipole approximation is valid. An initial test of its validity was performed by Foster and Krusch, who compared numerically generated skyrmion dynamics to the dipole approximation when the skyrmions are not spinning and are fixed in the attractive channel \cite{FS2014}. Their results indicate that the dipole approximation is reliable at large separations, but unreliable at small separations of the order 1fm.  They don't estimate the separation at which the dipole approximation ceases to be reliable.

%Hence in theory we can extend our graphs to this range of $r$. However, the inequalities tell us when the perturbation theory is valid and when the lagrangian \eqref{skyrmion-skyrmion lagrangian} still describes dipole dynamics; not when the dipole approximation for skyrmions is valid. An initial test of its validity was performed by Foster and Krusch, who compared numerically generated skyrmion dynamics to the dipole approximation when the skyrmions are not spinning and are fixed in the attractive channel \cite{FS2014}. We expect the approximation to break down before \eqref{ineqbreak} is reached.

\begin{figure}
\centering
	\includegraphics[width=0.8\textwidth]{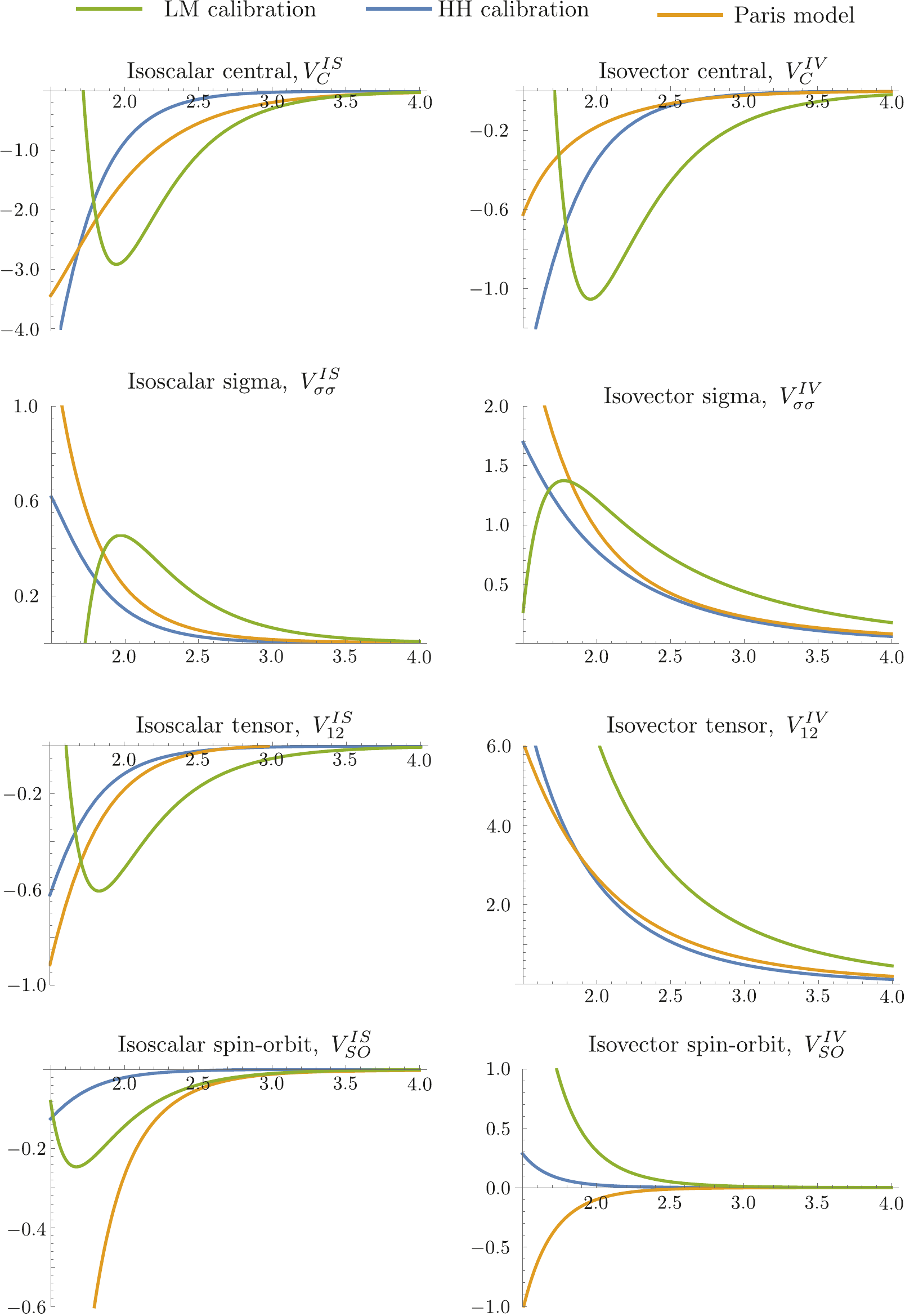}
	\caption{A comparison between the potentials generated from our calculation and the phenomenological Paris potential. All are plots of the potential (MeV) against separation $r$ (fm). }
	\label{fig:results}
\end{figure}

We plot the eight potentials from \eqref{NN potential} for the LM calibration, the HH calibration and from the Paris model in Figure \ref{fig:results}. For the long-range part of the interaction ($r \gtrsim 2 \text{ fm}$), both of the calibrations produce seven potentials with the correct sign. In the HH calibration six potentials closely match the Paris potentials even at intermediate separation. The LM calibration fails at shorter range. In both cases, the isoscalar spin-orbit force has the correct sign, though is too small in the HH calibration. The only major disagreement is with the isovector spin-orbit potential. This was successfully described in early works from the Skyrme model \cite{RN1987}, so perhaps nonlinear effects will resolve the disagreement.

Let us compare our results with earlier calculations  of the nucleon-nucleon potential from the Skyrme model.  In our calculation the potentials are expressed in terms of the expansion parameters $\rho$ and $M^{-1}$. At first order in $\rho$ only the isovector tensor and sigma potentials are non-zero. Hence second order perturbation theory was needed to generate non-trivial results.  In calculations done using first order perturbation theory and the product approximation \cite{NR1986,RD1988,Otofuji1988}, all of the potentials are non-zero at first order in $\rho$. In these calculations, all four isoscalar potentials had an incorrect sign -- we believe this is a failing of the product approximation. Our results, where only one isovector potential has an incorrect sign, are a substantial improvement on those calculations.

A direct comparison with the results of \cite{WAH1992,WA1992} is difficult, as those papers only compute potentials in particular channels and do not compute $V_C^{IS}$ etc. The method of \cite{WAH1992,WA1992} could in principle be used to compute six of the eight potentials in \eqref{NN potential}, but not the two spin-orbit potentials.  In this sense our method is more powerful. Another advantage of our approach over earlier methods is that it gives an explicit formula for the nucleon-nucleon interaction, but the price paid for this is that the formula doesn't capture short-range effects.

We also attempted to find a calibration of our model with $m_\pi=0$, by varying only $F_\pi$ and $e$ only. Here, it was much more difficult to find agreement between our model and the Paris potential. In particular, the isovector sigma potential $V^{IV}_{\sigma\sigma}$ had the wrong sign for all parameter choices that we tried. This suggests that a non-zero pion mass is an essential ingredient for producing realistic nucleon-nucleon interactions.

The calculation that we have presented was based on two key ingredients: perturbation theory beyond first order, and the inclusion of kinetic terms in the skyrmion-skyrmion interaction.  We now consider how both of these contribute to our final result.

It is clear from the results in subsection \ref{sec:massless pions} and \ref{app:pion mass} that if we had kept only first order terms and neglected terms of order $\rho^2$, we would be left with just the most basic long-range part of the nucleon-nucleon interaction, namely the one-pion-exchange potential.  Similarly, if we had retained terms of order $\rho^2$ but neglected terms of order $\rho^2/M$ our potential would not have contained a spin-orbit interaction.  Therefore it was necessary for us to work to order $\rho^2/M$ in order to have a complete description of the nucleon-nucleon potential.  These statements only apply to the dipole approximation of skyrmions; a different approximation could have produced a spin-orbit potential at lower orders. For example, models based on the product approximation gave a spin-orbit potential at first order in perturbation theory (but with the wrong sign) \cite{NR1987, RD1988,Otofuji1988}.

\begin{figure}
\centering
	\includegraphics[width=0.8\textwidth]{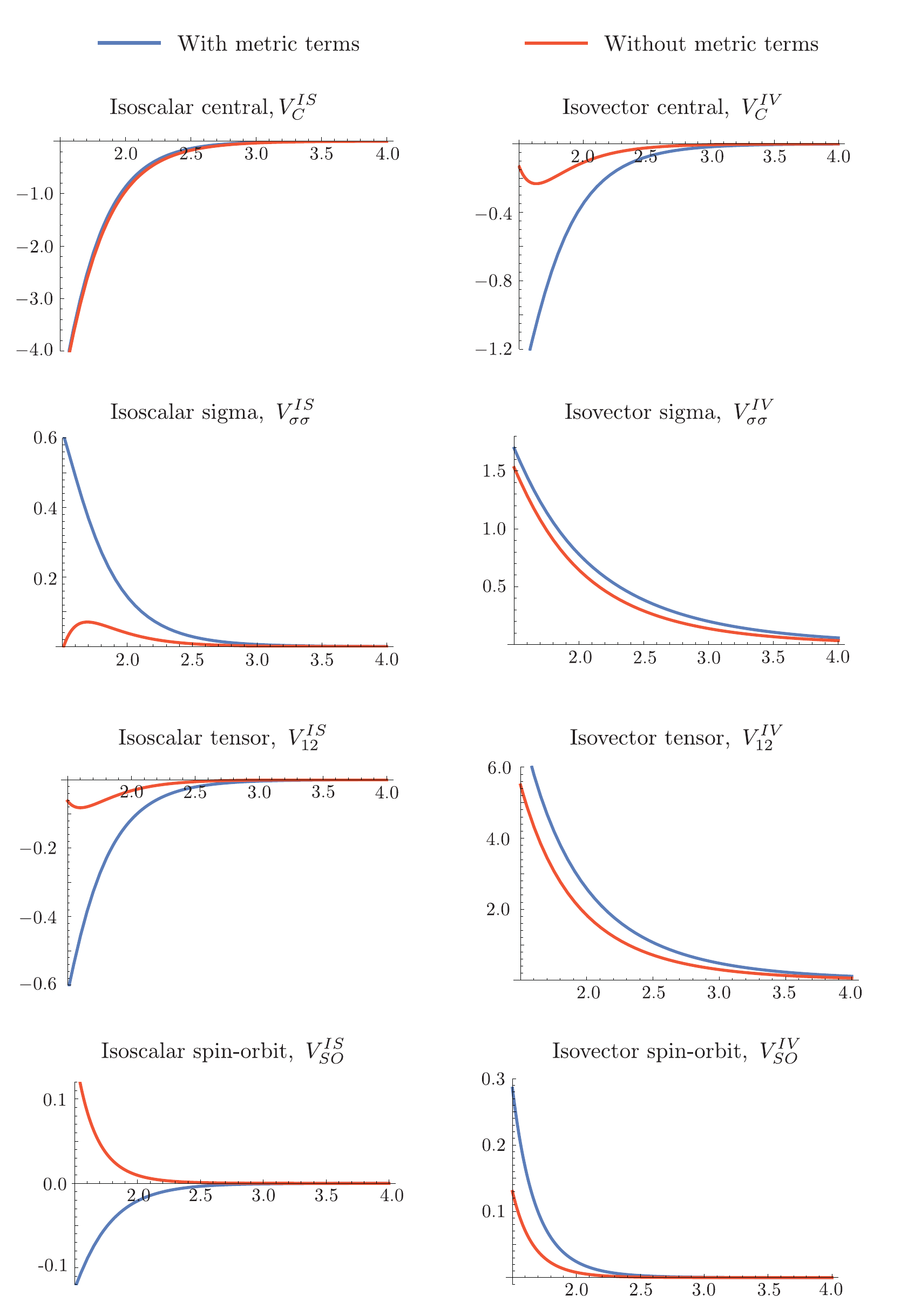}
	\caption{A comparison between the potentials when we do the calculation with and without the metric terms from \eqref{skyrmion-skyrmion lagrangian}. Both calculation are done using the HH calibration. All are plots of the potential (MeV) against separation $r$ (fm). }
	\label{fig:metpot}
\end{figure}

To investigate the influence of the kinetic terms in \eqref{skyrmion-skyrmion lagrangian} on the final result, we have re-done the calculation with the coefficients $A,B,C$ set to zero.  The results are presented in Figure \ref{fig:metpot}.  This makes clear that $V_{\sigma\sigma}^{IV}$, $V_{12}^{IS}$ and $V_{C}^{IS}$ receive their dominant contribution from the potential term (i.e.\ the term ``$D$'' in \eqref{skyrmion-skyrmion lagrangian}), but the other potentials receive significant contributions involving the kinetic terms. When kinetic terms are neglected, the isoscalar spin-orbit potential has the wrong sign. To obtain a realistic interaction, it appears essential to include the kinetic terms.

\section{Conclusions}

In summary, we have derived a nucleon-nucleon interaction from the Skyrme model using a method recently introduced in \cite{HH2020}.  Compared with earlier attempts based on the Skyrme model, we obtain a very good match with the long-range parts of the Paris potential.  Overall, these results provide an excellent starting point for describing the nucleon-nucleon interaction from the Skyrme model. Importantly, we can describe many features of the nucleon-nucleon interaction using a purely pionic theory.

It is interesting to compare our calculation with an earlier quantum mechanical study of the two-skyrmion system \cite{LMS1995}.  This paper studied bound states of two skyrmions, and in particular was successful in modelling the deuteron.  This was achieved using the Atiyah--Manton approximation \cite{AM1992}, which is able to describe skyrmions at both short and wide separations.  In contrast, the dipole interactions that we use are applicable only to well-separated skyrmions.  A shortcoming of the paper \cite{LMS1995} is that it only includes 10 degrees of freedom for the two-skyrmion system, whereas our approach includes 12.  In order to describe the spin and isospin states of well-separated skyrmions, 12 degrees of freedom are needed, so it is not possible to derive a conventional nucleon-nucleon potential from the configuration space in \cite{LMS1995}.  Therefore, a promising way to extend our results to intermediate and short separations would be to combine our result with \cite{LMS1995}: in other words, apply perturbation theory to quantum mechanics on a 12-dimensional configuration space of skyrmions obtained using the Atiyah--Manton approximation \cite{AM1992}.

%Our calculation is based on the dipole approximation to skyrmion dynamics, so does not capture the short-range part of the nucleon-nucleon interaction.  Hence to reach smaller values of $r$, we must first improve this approximation. The most complete description of the two-skyrmion configuration space is by Atiyah and Manton \cite{AM1992}, who used instantons to generate Skyrme fields. This gives a two-skyrmion configuration space which includes the minimal energy toroidal skyrmion as well as two widely separated skyrmions. Repeating the calculation presented here using the Atiyah-Manton approximation would allow us to study the intermediate- and short-range parts of the interaction.

The results presented in Figure \ref{fig:results} are promising, but to seriously judge the success of our calculation, we should compare directly with experimental data. This requires the calculation of phase shifts from our model, found by solving a Schr\"odinger equation based on the potentials. However, as explained above, we do not understand the potentials for small $r$ and these are needed for the calculation. Walet calculated phase shifts by imposing hard-core boundary conditions at $r=1$ \cite{Walet1993}. An advantage of the Skyrme model is that we should not require a hard-core: the geometry of the configuration space does not allow the skyrmions to get too close. It would be preferable to incorporate this fact into any future calculation.

The results suggest that the Skyrme model may provide an understanding of the nucleon-nucleon interaction using only pions. This is in contrast to the successful one-boson-exchange models which suggest that $\epsilon$-, $\Omega$- and $\rho$-mesons must be included. These can also be included in the Skyrme model \cite{Abada1996_2, GS2020, SN2018}, and it would be interesting to see their effect on the results presented here. To proceed, one must first understand the classical asymptotic interaction of skyrmions in models coupled to mesons, generalising the results of \cite{schroers1993,GP1994}. In fact, our approach could be adapted to any model which treats nuclei as quantised solitons. This includes holographic QCD, where nuclei are described as instantons on a curved spacetime \cite{SS2004}.

Finally, there are many modified Skyrme models. Authors have included different pionic terms \cite{ANSGW2013} and used modified potentials \cite{GHS2015,Gud2018} in the Skyrme lagrangian. Each modification will alter the results in Figure \ref{fig:results}. Since our calculation yields an explicit formula for the nucleon-nucleon interaction, it would be very easy to test these modified models by comparing their predictions for the nucleon-nucleon interaction.  We believe this new test will provide valuable insights for Skyrme phenomenology and help find the Skyrme model which best describes the physics of atomic nuclei.

\section*{Acknowledgments}

CJH is supported by The Leverhulme Trust as an Early Career Fellow.

\appendix

\section{The Clebsch-Gordon matrices $\kappa^{mn}_j$}
\label{app:kappa}

In this appendix we present the matrices $\kappa_j^{mn}$ that were used in our calculations.  We choose conventions such that the action of the spin operators $S^j$ on $\mathcal{H}_n\cong \CC^n\otimes\CC^n$ is given by $S^{j,nn}\otimes 1_n$, where
\begin{equation*}
S_1^{22}=\frac12\sigma_1=\frac12\begin{pmatrix}0&1\\1&0\end{pmatrix},\,
S_2^{22}=\frac12\sigma_2=\frac12\begin{pmatrix}0&-\ii\\\ii&0\end{pmatrix},\,
S_3^{22}=\frac12\sigma_3=\frac12\begin{pmatrix}1&0\\ 0&-1\end{pmatrix}
\end{equation*}
and
\begin{align*}
S_1^{44}&=\frac12 \begin{pmatrix}0&\sqrt{3}&0&0\\\sqrt{3}&0&2&0\\0&2&0&\sqrt{3}\\0&0&\sqrt{3}&0 \end{pmatrix}\\
S_2^{44}&=\frac12 \begin{pmatrix}0&-\ii\sqrt{3}&0&0\\\ii\sqrt{3}&0&-2\ii&0\\0&2\ii&0&-\ii\sqrt{3}\\0&0&\ii\sqrt{3}&0 \end{pmatrix}\\
S_3^{44}&=\frac12 \begin{pmatrix}3&0&0&0\\0&1&0&0\\0&0&-1&0\\0&0&0&-3 \end{pmatrix}.
\end{align*}
Then
\begin{align*}
\kappa^{22}_1 &= \frac{1}{\sqrt{3}}\begin{pmatrix}0&-1\\-1&0\end{pmatrix} &
\kappa^{22}_2 &= \frac{1}{\sqrt{3}}\begin{pmatrix}0&\ii\\-\ii&0\end{pmatrix} &
\kappa^{22}_3 &= \frac{1}{\sqrt{3}}\begin{pmatrix}-1&0\\ 0&1\end{pmatrix} \\
\kappa^{42}_1 &= \frac{1}{\sqrt{6}}\begin{pmatrix}-\sqrt{3}&0\\0&-1\\1&0\\0&\sqrt{3}\end{pmatrix} &
\kappa^{42}_2 &= \frac{1}{\sqrt{6}}\begin{pmatrix}\ii\sqrt{3}&0\\0&\ii\\\ii&0\\0&\ii\sqrt{3}\end{pmatrix} &
\kappa^{42}_3 &= \frac{1}{\sqrt{6}}\begin{pmatrix}0&0\\2&0\\0&2\\0&0\end{pmatrix}
\end{align*}
\begin{align*}
\kappa^{24}_1 &= \frac{1}{\sqrt{12}}\begin{pmatrix}\sqrt{3}&0&-1&0\\0&1&0&-\sqrt{3}\end{pmatrix} \\
\kappa^{24}_2 &= \frac{1}{\sqrt{12}}\begin{pmatrix}\ii\sqrt{3}&0&\ii&0\\0&\ii&0&\ii\sqrt{3}\end{pmatrix} \\
\kappa^{24}_3 &= \frac{1}{\sqrt{12}}\begin{pmatrix}0&-2&0&0\\0&0&-2&0\end{pmatrix} \\
%\end{align*}
%\begin{align*}
\kappa^{44}_1 &= \frac{1}{\sqrt{15}}\begin{pmatrix}0&-\sqrt{3}&0&0\\-\sqrt{3}&0&-2&0\\0&-2&0&-\sqrt{3}\\0&0&-\sqrt{3}&0\end{pmatrix} \\
\kappa^{44}_2 &= \frac{1}{\sqrt{15}}\begin{pmatrix}0&\ii\sqrt{3}&0&0\\-\ii\sqrt{3}&0&2\ii&0\\0&-2\ii&0&\ii\sqrt{3}\\0&0&-\ii\sqrt{3}&0\end{pmatrix} \\
\kappa^{44}_3 &= \frac{1}{\sqrt{15}}\begin{pmatrix}-3&0&0&0\\0&-1&0&0\\0&0&1&0\\0&0&0&3\end{pmatrix}.
\end{align*}

\section{The Laplace-Beltrami operator}
\label{app:beltrami}

In this appendix we derive the equation \eqref{LB definition} for the Laplace Beltrami operator.  Let $e^\mu$ be a frame for the cotangent bundle of a manifold and let $E_\mu$ be the dual frame for the tangent bundle.  Let $f_{\mu\nu}^\lambda$ be (locally defined) functions such that $[E_\mu,E_\nu]=f_{\mu\nu}^\lambda E_\lambda$.  Suppose that the metric is given by
\begin{equation}
g=g_{\mu\nu}e^\mu e^\nu.
\end{equation}
The standard definition for the Laplace-Beltrami operator acting on a function $\psi$ is
\begin{equation}
\triangle_g \psi = -\ast \dd\ast \dd \psi,
\end{equation}
in which $\ast$ denotes the Hodge star operator, defined by $u\wedge\ast v=g(u,v)\sqrt{g}e^1\wedge\ldots\wedge e^n$.  One finds that
\begin{equation}
\ast e^\nu = \sqrt{g}g^{\mu\nu}\iota_{E_\mu}\lrcorner (e^1\wedge\ldots\wedge e^n),
\end{equation}
in which $\iota$ denotes the interior product (so that $\iota_{E_1}e^1\wedge e^2\wedge\ldots\wedge e^n=e^2\wedge\ldots\wedge e^n$, $\iota_{E_2}e^1\wedge e^2\wedge\ldots\wedge e^n=-e^1\wedge e^3\ldots\wedge e^n$ etc.).  Therefore
\begin{equation}
-\ast \dd \psi = -\sqrt{g}g^{\mu\nu}(E_\nu\psi) \iota_{E_\mu} (e^1\wedge\ldots\wedge e^n)\label{eq:a1}.
\end{equation}

In order to evaluate $\dd\ast\dd\psi$ we need to to evaluate $\dd\iota_{E_\mu} (e^1\wedge\ldots\wedge e^n)$.  By the Cartan structure equations,
\begin{align}
\dd\iota_{E_\mu} (e^1\wedge\ldots\wedge e^n) &= -\iota_{E_\mu} \dd (e^1\wedge\ldots\wedge e^n) + \mathcal{L}_{E_\mu} (e^1\wedge\ldots\wedge e^n) \\
&= 0 + (\mathcal{L}_{E_\mu}e^1) \wedge e^2 \wedge \ldots\wedge e^n + e^1\wedge (\mathcal{L}_{E_\mu} e^2)\wedge \ldots\wedge e^n + \ldots \label{eq:a2},
\end{align}
with $\mathcal{L}$ denoting Lie derivative.  Now
\begin{equation}
\iota_{E_\lambda}(\mathcal{L}_{E_\mu} e^\nu)=\mathcal{L}_{E_\mu}(\iota_{E_\lambda}e^\nu) - \iota_{[E_\mu,E_\lambda]}e^\nu = 0 - f_{\mu\lambda}^\nu,
\end{equation}
so
\begin{equation}
\mathcal{L}_{E_\mu} e^\nu = - f_{\mu\lambda}^\nu E_\nu.
\end{equation}
Inserting this into equation \eqref{eq:a2} gives
\begin{align}
\dd\iota_{E_\mu} (e^1\wedge\ldots\wedge e^n) &= - (f_{\mu\lambda}^1e^\lambda)\wedge e^2\wedge\ldots \wedge e^n - e^1\wedge(-f_{\mu\lambda}^2e^\lambda)\wedge e^3\wedge\ldots\wedge e^n -\dots \\
&=-f_{\mu\lambda}^\lambda e^1\wedge\ldots\wedge e^n\label{eq:a3}.
\end{align}

Combining equations \eqref{eq:a1} and \eqref{eq:a3} gives
\begin{equation}
-\dd\ast\dd\psi = \sqrt{g}g^{\mu\nu}(E_\nu\psi) f_{\mu\lambda}^\lambda e^1\wedge\ldots\wedge e^n
-E_\lambda(\sqrt{g}g^{\mu\nu}(E_\nu\psi))e^\lambda\wedge \iota_{E_\mu} (e^1\wedge\ldots\wedge e^n)
\end{equation}
and thus
\begin{equation}
-\ast\dd\ast\dd\psi = -E_\mu(\sqrt{g}g^{\mu\nu}(E_\nu\psi)) + g^{\mu\nu}f_{\mu\lambda}^\lambda(E_\nu\psi),
\end{equation}
as claimed.

\section{Degenerate perturbation theory}
\label{app:perturbation theory}

In this appendix we derive equation \eqref{EH definition} for an effective hamiltonian using perturbation theory.  Let $H_F$ be a hamiltonian with eigenvalues $E_0<E_1<E_2<\ldots$.  Let $\ket{\psi^\alpha_0}$ be an orthonormal basis for the $E_0$-eigenspace.  Consider a deformation of $H_F$ of the form
\begin{equation}
H_F+\epsilon H_I.
\end{equation}
We seek deformed basis vectors $\ket{\psi^\alpha(\epsilon)}$ whose span is invariant under $H_F+\epsilon H_I$, and such that
\begin{equation}
\label{psi0}
\ket{\psi^\alpha(0)}=\ket{\psi^\alpha_0}.
\end{equation}
In other words, we require that
\begin{equation}
\label{Halphabeta definition}
(H_F+\epsilon H_I)\ket{\psi^\beta(\epsilon)} = \ket{\psi^\alpha(\epsilon)}H^{\alpha\beta}(\epsilon)
\end{equation}
for some $\epsilon$-dependent matrix $H^{\alpha\beta}$.  We also require these vectors, like $\ket{\psi^\alpha_0}$, to be orthonormal:
\begin{equation}
\label{psi orthonormal}
\braket{\psi^\alpha(\epsilon)}{\psi^\beta(\epsilon)}=\delta^{\alpha\beta}.
\end{equation}
In this situation $H^{\alpha\beta}(\epsilon)$ is the hermitian matrix of $H_F+\epsilon H_I$ acting on the subspace spanned by $\ket{\psi^\alpha(\epsilon)}$.  We can regard $H^{\alpha\beta}(\epsilon)$ as an effective hamiltonian describing the lowest eigenvalues of $H_F+\epsilon H_I$.

To calculate this effective hamiltonian one must solve the system \eqref{psi0}, \eqref{Halphabeta definition}, \eqref{psi orthonormal}.  This system does not have a unique solution, as one can make the replacement $\ket{\psi^\beta(\epsilon)}\to \ket{\psi^\alpha(\epsilon)}U^{\alpha\beta}(\epsilon)$, for any unitary matrix $U^{\alpha\beta}(\epsilon)$ and still have a solution.  In order to fix this degeneracy we impose the constraint
\begin{equation}
\label{psi constraint}
\braket{\psi^\alpha_0}{\psi^\beta(\epsilon)}=\braket{\psi^\alpha(\epsilon)}{\psi^\beta_0}.
\end{equation}

We seek to solve the system \eqref{psi0}, \eqref{Halphabeta definition}, \eqref{psi orthonormal}, \eqref{psi constraint} within the framework of perturbation theory.  That is, we seek a solution in the form
\begin{align}
\ket{\psi^\alpha(\epsilon)}&=\ket{\psi^\alpha_0}+\epsilon\ket{\psi^\alpha_1}+\epsilon^2\ket{\psi^\alpha_2}+\ldots \\
H^{\alpha\beta}(\epsilon)&=H^{\alpha\beta}_0 + \epsilon H^{\alpha\beta}_1+\epsilon^2H^{\alpha\beta}_2+\ldots
\end{align}
which formally solves the system to all orders in $\epsilon$.  In order to construct the solution we rewrite the equations in an iterative form.  Let $\Pi_N$ denote the projection onto the $E_N$-eigenspace of $H_F$.  Equations \eqref{psi orthonormal} and \eqref{psi constraint} imply that
\begin{equation}
\label{iteration 1}
\Pi_0 \ket{\psi^\alpha(\epsilon)}-\ket{\psi^\alpha_0} = -\frac{1}{2}\ket{\psi^\beta_0}\big(\bra{\psi^\beta(\epsilon)}-\bra{\psi^\beta_0}\big)\big(\ket{\psi^\alpha(\epsilon)}-\ket{\psi^\alpha_0}\big).
\end{equation}
Equations \eqref{Halphabeta definition} and \eqref{psi orthonormal} imply that
\begin{equation}
\label{iteration 2}
H^{\alpha\beta}(\epsilon) = \bra{\psi^\alpha(\epsilon)}(H_F+\epsilon H_I)\ket{\psi^\beta(\epsilon)}.
\end{equation}
Finally, equation \eqref{Halphabeta definition} implies that
\begin{equation}
\label{iteration 3}
\Pi_N\ket{\psi^\beta(\epsilon)}=\frac{1}{E_N-E_0}\left(\Pi_N\ket{\psi^\alpha(\epsilon)}\big(H^{\alpha\beta}(\epsilon)-E_0\delta^{\alpha\beta}\big)-\epsilon\Pi_NH_I\ket{\psi^\beta(\epsilon)}\right).
\end{equation}
for $N\neq0$.

Now we generate the perturbative solution using equations \eqref{iteration 1}, \eqref{iteration 2} and \eqref{iteration 3}.  First, equation \eqref{iteration 2} implies that
\begin{equation}
H^{\alpha\beta}_0 = \bra{\psi^\alpha_0}H_F\ket{\psi^\alpha_0}=E_0\delta^{\alpha\beta}.
\end{equation}
Equation \eqref{iteration 1} implies that
\begin{equation}
\Pi_0\ket{\psi_1^\alpha} = 0
\end{equation}
since the right hand side is $O(\epsilon^2)$.  Equation \eqref{iteration 2} implies that
\begin{equation}
H^{\alpha\beta}_1 = \bra{\psi^\alpha_0}H_I\ket{\psi^\beta_0}.
\end{equation}
Equation \eqref{iteration 3} implies that
\begin{equation}
\Pi_N\ket{\psi^\beta_1} = -\frac{1}{E_N-E_0}H_I^{N0} \ket{\psi^\beta_0},
\end{equation}
with $H_I^{MN}:=\Pi_M H_I\Pi_N$.  Thus altogether we have
\begin{equation}
\ket{\psi^\alpha_1} = - \sum_{N\neq 0}\frac{1}{E_N-E_0}H_I^{N0}\ket{\psi^\alpha_0}.
\end{equation}
This completes the solution to first order.  Now we compute the second order terms.  Equation \eqref{iteration 1} implies that
\begin{equation}
\Pi_0\ket{\psi^\alpha_2} = -\frac{1}{2}\sum_{N\neq0} \frac{1}{(E_N-E_0)^2}H_I^{0N}H_I^{N0}\ket{\psi^\alpha_0}.
\end{equation}
Equation \eqref{iteration 2} then implies that
\begin{equation}
H_2^{\alpha\beta} = - \sum_{N\neq 0}\frac{1}{E_N-E_0}\bra{\psi^\alpha_0}H_I^{0N}H_I^{N0}\ket{\psi^\beta_0}.
\end{equation}
Finally, equation \eqref{iteration 3} implies that
\begin{equation}
\Pi_N \ket{\psi^\alpha_2} = -\frac{1}{(E_N-E_0)^2}H_I^{N0}H_I^{00}\ket{\psi^\alpha_0} + \sum_{M\neq 0} \frac{1}{(E_N-E_0)(E_M-E_0)}H_I^{NM}H_I^{M0}\ket{\psi^\alpha_0},
\end{equation}
so that in total
\begin{multline}
\ket{\psi^\alpha_2} = -\frac{1}{2}\sum_{N\neq0} \frac{1}{(E_N-E_0)^2}H_I^{0N}H_I^{N0}\ket{\psi^\alpha_0} - \sum_{N\neq0}\frac{1}{(E_N-E_0)^2}H_I^{N0}H_I^{00}\ket{\psi^\alpha_0} \\
 + \sum_{M,N\neq0} \frac{1}{(E_N-E_0)(E_M-E_0)}H_I^{NM}H_I^{M0}\ket{\psi^\alpha_0}.
\end{multline}
This completes the solution to second order.  Finally, to third order equation \eqref{iteration 1} implies that
\begin{multline}
\Pi_0\ket{\psi^\alpha_3} =-\frac{1}{2}\sum_{N\neq0}\frac{1}{(E_N-E_0)^3}(H_I^{0N}H_I^{N0}H_I^{00}+H_I^{00}H_I^{0N}H_I^{N0})\ket{\psi^\alpha_0} \\
+\frac{1}{2}\sum_{M,N\neq0} \left(\frac{1}{(E_M-E_0)(E_N-E_0)^2}+\frac{1}{(E_M-E_0)^2(E_N-E_0)}\right)H_I^{0N}H_I^{NM}H_I^{M0}\ket{\psi^\alpha_0}
\end{multline}
and equation \eqref{iteration 2} implies that
\begin{multline}
H^{\alpha\beta}_3 = \sum_{M,N\neq0}\frac{1}{(E_N-E_0)(E_M-E_0)}\bra{\psi^\alpha_0}H_I^{0N}H_I^{NM}H_I^{M0}\ket{\psi^\beta_0} \\
- \frac{1}{2}\sum_{N\neq0} \frac{1}{(E_N-E_0)^2}\bra{\psi^\alpha_0}(H_I^{0N}H_I^{N0}H_I^{00}+H_I^{00}H_I^{0N}H_I^{N0})\ket{\psi^\beta_0}.
\end{multline}
Thus, our solution for $H^{\alpha\beta}$ is
\begin{multline}
H^{\alpha\beta} = \bigg\langle \psi^\alpha_0 \bigg| \bigg( H_F + \epsilon H_I^{00} - \epsilon^2\sum_{N\neq0} \frac{1}{E_N-E_0}H_I^{0N}H_I^{N0} \\
+\epsilon^3\sum_{M,N\neq0}\frac{1}{(E_N-E_0)(E_M-E_0)}H_I^{0N}H_I^{NM}H_I^{M0} \\
 - \frac{\epsilon^3}{2}\sum_{N\neq0} \frac{1}{(E_N-E_0)^2}(H_I^{0N}H_I^{N0}H_I^{00}+H_I^{00}H_I^{0N}H_I^{N0}) \bigg)\bigg|\psi^\beta_0\bigg\rangle + O(\epsilon^4).
\end{multline}

\section{Detailed calculation of potential with massless pions}
\label{app:massless pions}

In this section we presents the results of evaluating each of the terms in equations \eqref{second order} and \eqref{third order}.  We include these so that the reader can cross-check their calculations, should they wish to re-derive our result.
\begin{multline}
-4\rho^2\sum_{N>0}\frac{1}{E_N-E_0}D^{0N}D^{N0} \\
% &= -4\rho^2D_{ab}D_{cd}\sum_{N>0}\frac{1}{E_N-E_0}R_{ab}^{0N}R_{cd}^{N0} \\
= \frac{\rho^2\Lambda}{\hbar^2r^6}\left[-\frac{32}{9}-\frac{16}{27}\tautau + \frac{4}{27}(2+\tautau)(\sigsig-S_{12})\right]
\end{multline}
\begin{multline}
\frac{\rho^2\hbar^2}{\Lambda^2}\sum_{N>0} \frac{1}{E_N-E_0}\left(D^{0N}B^{N0}+B^{0N}D^{N0}\right)\\
%&=\frac{\rho^2\hbar^2}{\Lambda^2} D_{ab}B_{cd;ij}\sum_{N>0} \frac{1}{E_N-E_0}\left(R_{ab}^{0N}\left(S^1_iR_{cd}^{N0}S_2^j+S_2^jR_{cd}^{N0}S^1_i\right)+\left(S^1_iR_{cd}^{0N}S_2^j+S_2^jR_{cd}^{0N}S^1_i\right)R_{ab}^{N0}\right)\nonumber\\
=\frac{\rho^2}{\Lambda r^4}\left[-\frac{8}{27}-\frac{28}{81}\tautau+\frac{1}{81}(14+3\tautau)(\sigsig-2S_{12})\right].
\end{multline}
\begin{multline}
-\frac{\rho^2\hbar^4}{4\Lambda^4}\sum_{N>0} \frac{1}{E_N-E_0}B^{0N}B^{N0}\\
%&= -\frac{\rho^2\hbar^4}{4\Lambda^4}B_{ab;ij}B_{cd;kl}\sum_{N>0}\frac{1}{E_N-E_0}\left(S^1_iR_{ab}^{0N}S_2^j+S_2^jR_{ab}^{0N}S^1_i\right)\left(S^1_iR_{cd}^{N0}S_2^j+S_2^jR_{cd}^{N0}S^1_i\right) \nonumber\\
= \frac{\rho^2\hbar^2}{\Lambda^3 r^2}\left[-\frac{19}{54}-\frac{13}{324}\tautau-\frac{1}{1296}(26+17\tautau)(S_{12}+\sigsig)\right]
\end{multline}
\begin{multline}
-\frac{2\rho^2\hbar}{M\Lambda}\sum_{N>0} \frac{1}{E_N-E_0}\left(D^{0N}(P_iA_i+A_i^\dagger P_i)^{N0}+(P_iA_i+A_i^\dagger P_i)^{0N}D^{N0}\right) \\
%&=-\frac{\rho^2\hbar}{M\Lambda}\sum_{N>0} \frac{1}{E_N-E_0}\left\{P_i,D^{0N}(A_i+A_i^\dagger)^{N0}+(A_i+A_i^{\dagger})^{0N}D^{N0}\right\}\nonumber\\
%&\quad+\frac{\ii\rho^2\hbar^2}{M\Lambda}\sum_{N>0} \frac{1}{E_N-E_0}\left(D^{0N}\nabla_i(A_i-A_i^\dagger)^{N0}+\nabla_i(A_i-A_i^\dagger)^{0N}D^{N0}\right)\nonumber\\
%&\quad+\frac{\ii\rho^2\hbar^2}{M\Lambda}\sum_{N>0} \frac{1}{E_N-E_0}\left((A_i+A_i^\dagger)^{0N}\nabla_iD^{N0}-\nabla_iD^{0N}(A_i+A_i^\dagger)^{N0}\right)\nonumber\\
=\frac{\rho^2}{M r^6}\left[-\frac{32}{9}-\frac{40}{27}\tautau+\left(\sigsig-S_{12}\right)\left(\frac{20}{27}+\frac{2}{9}\tautau\right)\right]\\
+\frac{\rho^2}{M\hbar r^6}\left[-\frac{4}{3}+\frac{4}{9}\tautau\right]\Ls
\end{multline}
\begin{multline}
\frac{\rho^2\hbar^3}{2M\Lambda^3}\sum_{N>0} \frac{1}{E_N-E_0}\left(B^{0N}(P_iA_i+A_i^\dagger P_i)^{N0}+(P_iA_i+A_i^\dagger P_i)^{0N}B^{N0}\right) \\
%&= \frac{\rho^2\hbar^3}{4M\Lambda^3}\sum_{N>0} \frac{1}{E_N-E_0}\left\{P_i,B^{0N}(A_i+A_i^\dagger)^{N0}+(A_i+A_i^{\dagger})^{0N}B^{N0}\right\}\nonumber\\
%&\quad-\frac{\ii\rho^2\hbar^4}{4M\Lambda^3}\sum_{N>0} \frac{1}{E_N-E_0}\left(D^{0N}\nabla_i(A_i-A_i^\dagger)^{N0}+\nabla_i(A_i-A_i^\dagger)^{0N}D^{N0}\right)\nonumber\\
%&\quad-\frac{\ii\rho^2\hbar^3}{4M\Lambda^3}\sum_{N>0} \frac{1}{E_N-E_0}\left((A_i+A_i^\dagger)^{0N}\nabla_iD^{N0}-\nabla_iD^{0N}(A_i+A_i^\dagger)^{N0}\right)\nonumber\\
=\frac{\rho^2\hbar^2}{M\Lambda^2r^4}\left[-\frac{26}{27}+\frac{1}{9}\tautau+\left(\sigsig-2S_{12}\right)\left(\frac{43}{162}+\frac{61}{972}\tautau\right)\right]\\
+\frac{\rho^2\hbar}{M\Lambda^2r^4}\left[\frac{20}{27}+\frac{5}{81}\tautau\right]\Ls.
\end{multline}
\begin{multline}
\frac{4\ii\rho^2\hbar}{M}\sum_{N>0} \frac{1}{(E_N-E_0)^2}\big\{P_i,\, \nabla_iD^{0N}D^{N0}-D^{0N}\nabla_iD^{N0}\big\} \\
= \frac{\rho^2\Lambda^2}{M\hbar^3 r^8}\left(\frac{16}{27}+\frac{40}{81}\tautau\right)\Ls
\end{multline}
\begin{multline}
-\frac{\ii\rho^2\hbar^3}{M\Lambda^2}\sum_{N>0} \frac{1}{(E_N-E_0)^2}\big\{P_i,\, \nabla_iD_I^{0N}B_I^{N0}-D_I^{0N}\nabla_iB_I^{N0}+\nabla_iB_I^{0N}D_I^{N0}-B_I^{0N}\nabla_iD_I^{N0}\big\} \\
= \frac{\rho^2}{M\hbar r^6}\left(\frac{52}{81}+\frac{22}{243}\tautau\right)\Ls
\end{multline}
\begin{multline}
\frac{\ii\rho^2\hbar^5}{4M\Lambda^4}\sum_{N>0} \frac{1}{(E_N-E_0)^2}\big\{P_i,\, \nabla_iB_I^{0N}B_I^{N0}-B_I^{0N}\nabla_iB_I^{N0}\big\} \\
= \frac{\rho^2\hbar}{M\Lambda^2 r^4}\left(\frac{7}{972}+\frac{89}{5832}\tautau\right)\Ls
\end{multline}
\begin{multline}
\frac{4\rho^2\hbar^2}{M}\sum_{N>0} \frac{1}{(E_N-E_0)^2}\nabla_iD_I^{0N}\nabla_iD_I^{N0} \\
= \frac{\rho^2\Lambda^2}{M\hbar^2r^8}\left[ \frac{800}{27}+\frac{560}{81}\tautau + (4S_{12}-5\sigsig)\left(\frac{56}{81} + \frac{68}{243}\tautau\right) \right]
\end{multline}
\begin{multline}
-\frac{\rho^2\hbar^4}{M\Lambda^2}\sum_{N>0} \frac{1}{(E_N-E_0)^2}\big(\nabla_iD_I^{0N}\nabla_iB_I^{N0}+\nabla_iB_I^{0N}\nabla_iD_I^{N0}\big) \\
= \frac{\rho^2}{Mr^6}\left[ \frac{88}{27} + \frac{148}{81}\tautau + (S_{12}-\sigsig) \left(\frac{74}{81} + \frac{59}{243}\tautau\right)\right]
\end{multline}
\begin{multline}
\frac{\rho^2\hbar^6}{4M\Lambda^4}\sum_{N>0} \frac{1}{(E_N-E_0)^2}\nabla_iB_I^{0N}\nabla_iB_I^{N0} \\
= \frac{\rho^2\hbar^2}{M\Lambda^2 r^4}\left[ \frac{89}{162} + \frac{103}{972}\tautau + (2S_{12}-\sigsig)\left(\frac{103}{5832} + \frac{281}{34992}\tautau\right)\right]
\end{multline}

\section{Potentials with non-zero pion mass}
\label{app:pion mass}

Setting $s = m_\pi r / \hbar$, the isoscalar nucleon-nucleon potentials are given by
\begin{align}
	V_C^{IS} = &-\rho ^2\frac{ e^{-2 s}}{108 \Lambda ^3 r^6 \hbar ^2} \big(64 \Lambda ^4 \left(s^4+4 s^3+10 s^2+12 s+6\right) \nonumber\\&\quad+16 \Lambda ^2
	r^2 \hbar ^2 \left(s^3+2 s^2+4 s+2\right) -17 r^4 \hbar ^4 \left(s^2+2\right) 
	\big)\nonumber \\
	&+\frac{\rho ^2}{M}\frac{ e^{-2 s} }{486 \Lambda ^2 r^8 \hbar ^2} \big(160 \Lambda ^4 \left(s^6+6 s^5+27 s^4+84s^3+162 s^2+180 s+90\right)\nonumber\\&\quad+8 \Lambda ^2 r^2 \hbar ^2 \left(83 s^5-3 s^4-12 s^3-30 s^2-36 s-18\right)\nonumber\\&\quad+r^4 \hbar ^4 \left(103 s^4-270s^3+798 s^2+1704 s+852\right)\big)\\
	\nonumber \\
	V_{12}^{IS} =
	&-\rho ^2\frac{ e^{-2 s}}{648 \Lambda ^3 r^6 \hbar ^2} \big(64 \Lambda ^4 \left(s^3+4 s^2+6 s+3\right)\nonumber\\&\quad+56 \Lambda ^2 r^2 \hbar ^2 \left(2 s^2+5 s+4\right)
	+13 r^4 \hbar ^4 (s+1)\big)\nonumber \\
	&+\frac{\rho ^2}{M}\frac{ e^{-2 s}}{5832  \Lambda ^2 r^8 \hbar ^2} \big(448 \Lambda ^4 \left(s^5+8 s^4+30 s^3+63 s^2+72 s+36\right)\nonumber\\&\quad+56 \Lambda ^2 r^2 \hbar ^2 \left(26 s^4+71 s^3+76 s^2+36 s+18\right)\nonumber\\&\quad+r^4 \hbar ^4 \left(913 s^3-609 s^2-4052 s-3538\right)\big)\\
	\nonumber \\
	V_{\sigma\sigma}^{IS} = &\,\,\rho^2\frac{ e^{-2 s}}{648 \Lambda ^3 r^6 \hbar ^2}  \big(64 \Lambda ^4 \left(2 s^3+5 s^2+6 s+3\right)+112 \Lambda ^2 r^2 \hbar ^2 \left(2 s^2+2 s+1\right)\nonumber\\&\quad+13 r^4 \hbar ^4 (2s-1)\big)\nonumber\\ 
	&-\frac{\rho ^2}{M}\frac{ e^{-2 s}}{5832 \Lambda ^2 r^8 \hbar ^2} \big(448 \Lambda ^4 \left(2 s^5+13 s^4+42 s^3+81 s^2+90 s +45\right)\nonumber\\&\quad+112 \Lambda ^2 r^2 \hbar ^2 \left(26 s^4+32 s^3+28 s^2+18 s+9\right)\nonumber\\&\quad+r^4 \hbar ^4 \left(1826 s^3-3957 s^2-4510 s-2255\right)\big)\\
	\nonumber \\
	V_{SO}^{IS} =  \,\,&\frac{\rho ^2}{M}\frac{ e^{-2 s}}{972 \Lambda ^2 r^8 \hbar ^2} \big(64 \Lambda ^4 \left(s^2+3 s+3\right)^2\nonumber\\&\quad-32 \Lambda ^2 r^2 \hbar ^2 \left(16 s^3+37s^2+42 s+21\right)-r^4 \hbar ^4 \left(29 s^2+274 s+245\right)\big) \, .
\end{align}

The isovector potentials are given by
\begin{align}
	V_C^{IV} = &-\rho ^2\frac{ e^{-2 s}}{648 \Lambda ^3 r^6 \hbar ^2} \big(64 \Lambda ^4 \left(s^4+4 s^3+10 s^2+12 s+6\right)\nonumber\\&\quad+112 \Lambda ^2
	r^2 \hbar ^2 \left(s^3+2 s^2+4 s+2\right)+13 r^4 \hbar ^4 \left(s^2+2\right)\big)\nonumber \\
	&+\frac{\rho ^2}{M}\frac{ e^{-2 s}}{5832  \Lambda ^2 r^8 \hbar ^2} \big(448 \Lambda ^4 \left(s^6+6 s^5+27 s^4+84 s^3+162 s^2+180 s+90\right)\nonumber\\&\quad+16 \Lambda ^2 r^2 \hbar ^2 \left(91 s^5+57 s^4+156s^3+246 s^2+252 s+126\right)\nonumber\\&\quad+r^4 \hbar ^4 \left(913 s^4-2304 s^3-930 s^2-3300 s-1650\right)\big)\\
	V^{IV}_{12} =& \,\,\rho\frac{ e^{-s}}{27 \Lambda ^2 r^3} \left(2 \Lambda ^2 \left(s^2+3 s+3\right)+r^2 \hbar ^2 (s+1) \right)- \frac{\rho}{M}\frac{ \hbar ^2e^{-s} }{27 \Lambda r^3} \left(s^3-3 s^2-12 s-12\right) \nonumber  \\
	&+\rho^2\frac{e^{-2 s}}{1296 \Lambda ^3 r^6 \hbar ^2} \big(-64\Lambda ^4 \left(s^3+4 s^2+6 s+3\right)-24 \Lambda ^2 r^2 \hbar ^2 \left(2 s^2+5 s+4\right)\nonumber\\&\quad+19 r^4 \hbar ^4 (s +1)\big) 
	\nonumber\\ &+
	\frac{\rho^2}{M}\frac{e^{-2 s}}{34992 \Lambda ^2 r^8 \hbar ^2} \big(1088 \Lambda ^4 \left(s^5+8 s^4+30 s^3+63 s^2+72 s+36\right)\nonumber\\&\quad+8 \Lambda ^2 r^2 \hbar ^2 \left(514 s^4+1315 s^3+1148 s^2+180 s+90\right)\nonumber\\&\quad+r^4 \hbar ^4 \left(677 s^3+2355 s^2+3356 s+1678\right)\big)\\
	\nonumber \\
	V^{IV}_{\sigma\sigma} = &\,\,\rho\frac{ e^{-s}}{27 \Lambda ^2 r^3} \left(2 \Lambda ^2 s^2+ r^2\hbar ^2(s-2) \right) -  \frac{\rho}{M}\frac{ \hbar^2e^{-s}}{27 \Lambda r^3} s^2 (s-6) \nonumber\\&+ \rho^2\frac{ e^{-2 s}}{1296 \Lambda ^3 r^6 \hbar ^2} \big(64	\Lambda ^4 \left(2 s^3+5 s^2+6 s+3\right)+48 \Lambda ^2 r^2 \hbar ^2 \left(2 s^2+2 s+1\right)\nonumber\\&\quad+19 r^4 \hbar ^4 (1-2 s)\big) \nonumber \\ &-\frac{\rho^2}{M} \frac{ e^{-2 s}}{34992 \Lambda ^2 r^8 \hbar ^2} \big(1088 \Lambda ^4 \left(2 s^5+13 s^4+42 s^3+81 s^2+90 s+45\right)\nonumber\\&\quad+16 \Lambda ^2 r^2 \hbar ^2 \left(514 s^4+544 s^3+332 s^2+90s+45\right)\nonumber\\&\quad+r^4 \hbar ^4 \left(1354 s^3+2679 s^2+2650 s+1325\right)\big)\\
	\nonumber \\
	V_{SO}^{IV}=\,\,&\frac{\rho^2}{M}\frac{e^{-2 s}}{5832  \Lambda ^2 r^8 \hbar ^2} \big(320 \Lambda ^4 \left(s^2+3 s+3\right)^2\nonumber\\&\quad+16 \Lambda ^2 r^2 \hbar ^2 \left(29 s^3+224s^2+390 s+195\right)+r^4 \hbar ^4	\left(557 s^2+1330 s+773\right)\big) \, .
\end{align}

\bibliographystyle{ieeetr}
%\bibliography{nn.bib}

\end{document}